\documentclass[aps,prl,twocolumn,nofootinbib,balance,superscriptaddress,floats]{revtex4}
\usepackage[a4paper, total={7.2in, 10in}]{geometry}

\usepackage{latexsym}
\usepackage{dcolumn}
\usepackage{amsmath}
\usepackage{float}

\usepackage[pdftex]{graphicx}
\usepackage{epstopdf}
\usepackage{pdfpages}
\usepackage{soul}
\usepackage{upgreek}
\usepackage{tabularx}

\newcommand\blfootnote[1]{%
  \begingroup
  \renewcommand\thefootnote{}\footnote{#1}%
  \addtocounter{footnote}{-1}%
  \endgroup
}

\begin{document}
\raggedbottom
%%\preprint{APS/123-QED}

%\title{Electrically-driven Acousto-optic Modulators and Broadband Non-reciprocity in Silicon Photonics} % Force line breaks with \\

%\title{Electrically-driven Acousto-optics and Broadband Non-reciprocity in Silicon Photonics} 

\title{Electrically-driven Acousto-optics and Broadband Non-reciprocity in Silicon Photonics} 

\author{Eric A. Kittlaus$^{1,\dagger}$}
\noaffiliation
%\affiliation{Department of Applied Physics, Yale University, New Haven, CT 06520 USA.}
%\email{eric.kittlaus@jpl.nasa.gov}
\author{William M. Jones}
\affiliation{Jet Propulsion Laboratory, California Institute of Technology, Pasadena, CA 91109 USA.}
\author{Peter T. Rakich}
\affiliation{Department of Applied Physics, Yale University, New Haven, CT 06520 USA.} 
\author{Nils T. Otterstrom}
\affiliation{Department of Applied Physics, Yale University, New Haven, CT 06520 USA.}
\author{Richard E. Muller}
\affiliation{Jet Propulsion Laboratory, California Institute of Technology, Pasadena, CA 91109 USA.}
\author{Mina Rais-Zadeh}
\affiliation{Jet Propulsion Laboratory, California Institute of Technology, Pasadena, CA 91109 USA.}
%\author{Siamak Forouhar}
%\affiliation{Jet Propulsion Laboratory, California Institute of Technology, Pasadena, CA 91109 USA.}

%\author{Eric A. Kittlaus$^{1,\dagger}$, Nils T. Otterstrom$^{1}$, and Peter T. Rakich$^{1,\ddagger}$}
%\makeatother
%\affiliation{$^1$Department of Applied Physics, Yale University, New Haven, CT 06520 USA.}

\date{\today}

%Previous integrated isolators have not been able to achieve large nonreciprocity, large operating bandwidth, and low insertion losses. This one doesn't either, but it's closer than the others.

%As a result, there is a pressing need for linear nonreciprocal devices which can be monolithically integrated into photonic circuits. 

\begin{abstract}
Emerging technologies based on tailorable interactions between photons and phonons promise new capabilities ranging from high-fidelity microwave signal processing to non-reciprocal optics and quantum state control. While such light-sound couplings have been studied in a variety of physical systems, many implementations rely on non-standard materials and fabrication schemes that are challenging to co-implement with standard integrated photonic circuitry. Notably, despite significant advances in integrated electro-optic modulators, related acousto-optic modulator concepts have remained relatively unexplored in silicon photonics. In this article, we demonstrate direct acousto-optic modulation within silicon photonic waveguides using electrically-driven surface acoustic waves (SAWs). By co-integrating SAW transducers in piezoelectric aluminum nitride with a standard silicon-on-insulator photonic platform, we harness silicon's strong elasto-optic effect to mediate non-local light-sound coupling. Through lithographic design, acousto-optic phase modulators and single-sideband amplitude modulators in the range of 1-5 GHz are fabricated, exhibiting refractive index modulation strengths comparable to existing electro-optic technologies. Extending this traveling-wave, acousto-optic interaction to centimeter-scales, we create electrically-driven non-reciprocal modulators in silicon. Non-reciprocal operation bandwidths of $>$100 GHz and insertion losses $<$0.6 dB are achieved. Building on these results, we show that unity-efficiency non-reciprocal modulation, necessary to produce a robust acousto-optic isolator, is within reach. The acousto-optic modulator design is compatible with both complementary metal–oxide–semiconductor (CMOS) fabrication processes, and with existing silicon photonic device technologies. These results represent a promising new approach to implement compact and scalable acousto-optic modulators, frequency-shifters, and non-magnetic optical isolators and circulators in integrated photonic circuits.
\end{abstract}

\maketitle

\section{Introduction}
%kuhn1970deflection
%tsai1975high
%de2005modulation
%li2019electromechanical2
%fan2019spectrotemporal
%satzinger2018quantum,chu2018creation
Light-sound\blfootnote{$\dagger$ eric.kittlaus@jpl.nasa.gov} interactions in solid-state systems show strong potential as a basis for diverse applications ranging from quantum information control \protect{\cite{aspelmeyer2014cavity, verhagen2012quantum,andrews2014bidirectional}} to microwave and optical signal processing \protect{\cite{eichenfield2009optomechanical,safavi2011electromagnetically,fan2016integrated,bernier2017nonreciprocal,safavi2019controlling}}. Nonlinear optomechanical couplings have been harnessed for microwave filtering and synthesis \protect{\cite{tomes2009photonic,jiang2012high,fong2014microwave,marpaung}}, optical non-reciprocity \protect{\cite{Kang2011,Dong2015,kim2015,Shen2016,Ruesink2016,Kim2017}}, and for chip-based amplifiers \protect{\cite{pant,Kittlaus2017,otterstrom2019resonantly}} and lasers \protect{\cite{lee2012chemically,Morrison17,Otterstrom2017,gundavarapu2019sub}}. Meanwhile, linear acousto-optic interactions have been widely studied in piezoelectric material systems for optical modulation \protect{\cite{kuhn71,sasaki74,Ohmachi77,gorecki1997silicon,lima2006compact,Tadesse2014,tadesse2015acousto,fan2016integrated,balram2017acousto,liu2019electromechanical}} and opto-acoustic gating \protect{\cite{balram2017acousto}}, and have recently attracted significant attention as a potential mechanism for microwave-optical conversion \protect{\cite{jiang2019lithium,shao2019microwave,wu2020microwave}} and non-reciprocal routing \protect{\cite{Yu2009,sohn17,Kittlaus2018}}. 

In particular, distributed acousto-optic modulation has emerged as a promising means to implement low-loss, non-magnetic isolators and circulators in integrated photonic circuits \protect{\cite{Kang2011,Yu2009,sohn17,Kittlaus2018,sohn2019direction}}. Such modulation-based approaches are particularly attractive for chip-scale implementation since they avoid considerable fabrication challenges and excess optical absorption associated with miniaturized magneto-optic isolators \protect{\cite{Shoji2008,Bi2011,huang16,Sounas2017}}. Recent device demonstrations have produced non-reciprocal acousto-optic modulation \protect{\cite{sohn17}} and broadband operation \protect{\cite{Kittlaus2018}}, but have relied on suspended optomechanical waveguides, and either narrowband optical resonators \protect{\cite{sohn17}} or optical pumping \protect{\cite{Kittlaus2018}} to achieve the non-reciprocal effect, posing potential challenges to system scalability. Furthermore, improvements in modulation strength are necessary to translate these nascent technologies into robust non-reciprocal components. Nonetheless, such advances represent immense potential; high-performance, reconfigurable integrated isolators and circulators are feasible if strong acousto-optic interactions can be implemented on-chip.

%Nonetheless, the potential impact of such advances is immense; high-performance, reconfigurable integrated isolators and circulators should be possible with the creation of strong acousto-optic interactions on-chip.
%These impressive demonstrations have been realized in a variety of different systems, including high-$Q$ resonators, piezoelectric AlN or LiNbO$_3$, and in bulk geometries.

However, acousto-optic devices have remained largely unrealized in silicon---the leading material for integrated photonics. Despite an impressive array of silicon photonic device technologies, including miniaturized electro-optic modulators \protect{\cite{Xu2005,Chen09,reed2010silicon,watts2010low,tbj2012,weigel2018bonded,wang2018integrated}}, detectors \protect{\cite{Chen09,tbj2012}}, signal processors \protect{\cite{sun2015single,perez2017multipurpose}}, and oscillators \protect{\cite{rong2005,Otterstrom2017,huang2019high}}, silicon's lack of piezoelectric coupling, and intrinsic anti-guiding of acoustic waves in silicon-on-insulator have stifled progress in silicon-based acousto-optics. Related nonlinear couplings have been realized via on-chip stimulated Brillouin scattering \protect{\cite{shinnatcomm,Kittlaus2016,vanlaernatphoton,Otterstrom2017}}, enabled by silicon's exceptionally strong elasto-optic effect. These devices have thus far used relatively inefficient nonlinear optical transduction of acoustic waves within non-standard suspended \protect{\cite{shinnatcomm,Kittlaus2016,vanlaernatphoton}} or slot waveguide structures \protect{\cite{sarabalis2017release}}. Attempts to implement direct elasto-optic modulation in silicon through capacitive coupling \protect{\cite{van2018electrical}}, or photo-acoustic effects \protect{\cite{munk2019surface}} have shown potential for powerful degrees of control. However, at present, modest $(<10^{-3})$ acousto-optic scattering efficiencies in these systems pose a limitation to their practical utility.  

In this article, we demonstrate electrically-driven acousto-optic modulation in silicon photonic waveguides and use this capability to create broadband, integrated non-reciprocal optical modulators. Direct electromechanical transduction is achieved using a CMOS-compatible aluminum nitride on silicon-on-insulator (AlN-on-SOI) material platform. Surface acoustic waves (SAWs) launched at integrated, low-loss optical waveguides produce linear acousto-optic modulation via traveling-wave elasto-optic coupling. Through lithographic control of device geometry, we realize both optical phase modulation and single-sideband amplitude modulation in the frequency range of 1-5 GHz. These modulators utilize no suspended structures or optical pumping, and achieve modulation figures of merit $V_\pi L = $ 1$-$3 V$\cdot$cm, comparable to existing electro-optic modulator technologies. Next, we show how these interactions can be extended to longer interaction lengths, a necessary condition for the creation of non-reciprocal devices based on optical modulation. Using a serpentine waveguide structure, we demonstrate enhanced modulation efficiency ($>10^{-1}$) and electrically-driven non-reciprocal modulation and optical mode conversion over a 100 GHz (0.8 nm) optical bandwidth, representing a significant step toward practical optical isolators and circulators in silicon. The implementation of robust acousto-optic modulators on-chip may enable a range of applications including chip-based heterodyne detection, frequency modulation, waveform synthesis, and switching. More generally, this design approach opens the door to a variety of electro-opto-mechanical devices, as well as flexible acousto-optic modulators and non-reciprocal operations, that can be directly integrated within silicon photonic circuits.

\begin{figure*}[t]
\centering%\vspace{-10pt}
\includegraphics[width=0.8\linewidth]{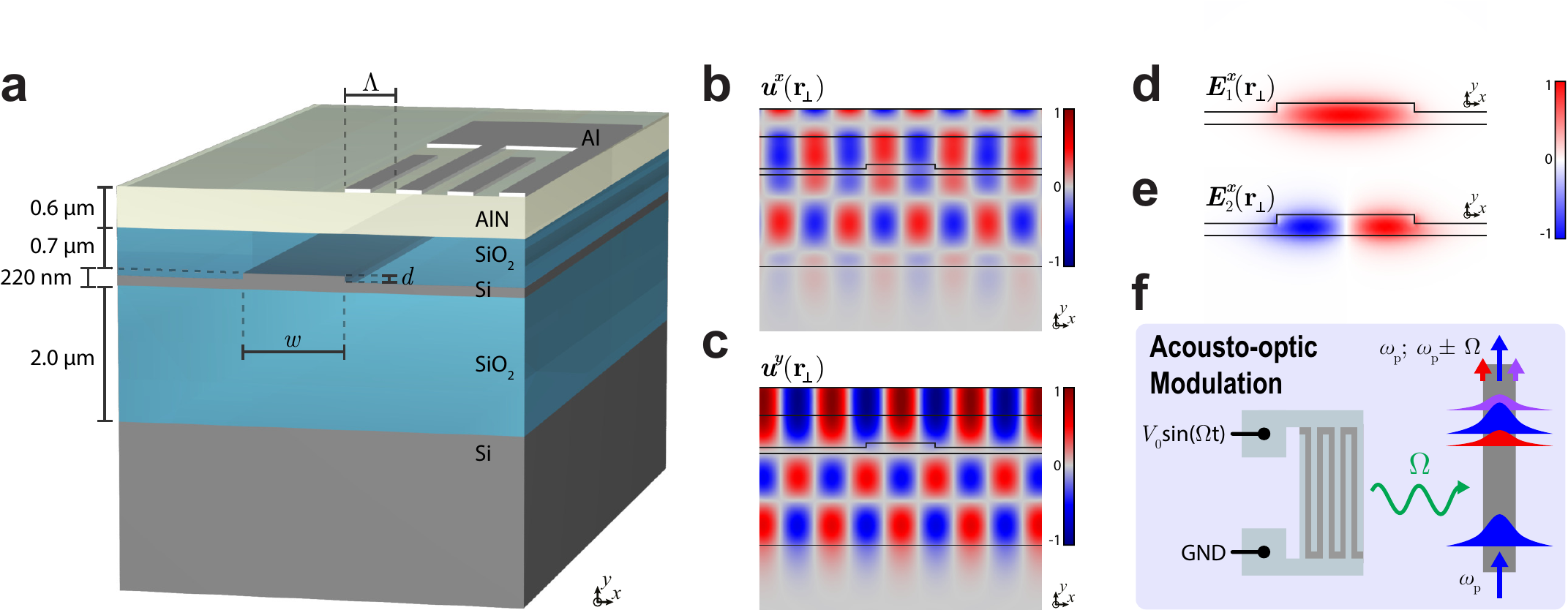}
\caption{Integrated acousto-optic platform in silicon photonics. (a) Artistic representation of the device cross-section; a silicon ridge waveguide of width $w$ and etch depth $d$ is clad in silicon dioxide, and lies beneath a top layer of piezoelectric aluminum nitride. Surface acoustic wave transducers consisting of inter-digitated electrodes with pitch $\Lambda$ are patterned in aluminum. (b) and (c) depict simulated $x-$ and $y-$displacement profiles of one surface acoustic wave around a wavelength of $\Lambda_\textup{ac} = 1.5$ $\upmu$m (frequency $\Omega/2\pi$ = 3.1 GHz). Note that the elastic wave extends into the optical waveguide layer. (d) and (e) depict the $x-$directed electric field profiles of symmetric and anti-symmetric optical modes guided by the silicon ridge waveguide. (f) shows the basic principle of the acousto-optic modulation process; a time-harmonic voltage with frequency $\Omega$ applied to an inter-digitated electrode transduces a surface acoustic wave at frequency $\Omega$, which then modulates light (frequency $\omega_\textup{p}$) guided in an optical waveguide through the elasto-optic effect. This modulation process produces optical energy transfer to frequency-detuned sidebands at $\omega_\textup{p}\pm\Omega$.}
\label{fig:intro}
\end{figure*}

\section{Results}

\subsection{Acousto-optic Device Platform}

%The mechanism of the inter-modal emit-receive process is diagrammed in Fig 1a. Strong pump waves at frequencies $\omega_{1,p}$ and $\omega_{1,s}$ are injected into distinct symmetric and anti-symmetric optical modes of an optical waveguide. These optical fields excite a travelling-wave acoustic phonon at difference frequency $\Omega = \omega_{1,p} - \omega_{1,s}$ through a stimulated inter-modal Brillouin scattering (SIMS) process. A separate optical 

We fabricate on-chip acousto-optic modulators (AOMs) using the device platform depicted in Fig. \ref{fig:intro}a. Optical ridge waveguides are fabricated on a standard silicon-on-insulator wafer, with buried oxide and device layer thicknesses of 2 $\upmu$m and 220 nm, respectively. The etch depth of the ridge waveguides $d$ = 90 nm, and the waveguide width $w$ is lithographically defined to values between 450 and 1500 nm depending on the specific device design. A 700 nm silicon dioxide overcladding is deposited by plasma-enhanced chemical vapor deposition (PECVD), and the wafer is annealed at 750$^\circ$ C in nitrogen gas to remove impurities in the PECVD silicon dioxide via diffusion. 

Polycrystalline aluminum nitride (AlN) is deposited on top of the oxide layer using a magnetron reactive sputtering process. AlN was chosen for the top clad material because it offers strong piezoelectric coupling, can be deposited c-axis-oriented with good film uniformity (X-ray diffraction rocking curve full-width at half-maximum (FWHM) = 1.6$^\circ$), and because its low-temperature sputter deposition process is CMOS-compatible \protect{\cite{xiong2012aluminum}}. Sputtered AlN can also be deposited on a variety of other substrates, including Si and Si$_{3}$N$_{4}$, making this design approach applicable to other optical device platforms. The film thickness of 600 nm was chosen according to finite-element simulations to optimize acousto-optic device performance (see Supplementary Note II). Electrical contact pads and inter-digitated surface acoustic wave transducers (IDTs) consisting of equally-spaced fingers with inter-digital pitch $\Lambda$ are patterned in 50 nm-thick aluminum on top of the piezoelectric AlN using a liftoff process. 

\begin{figure*}[htbp]
\centering%\vspace{-10pt}
\includegraphics[width=.66\linewidth]{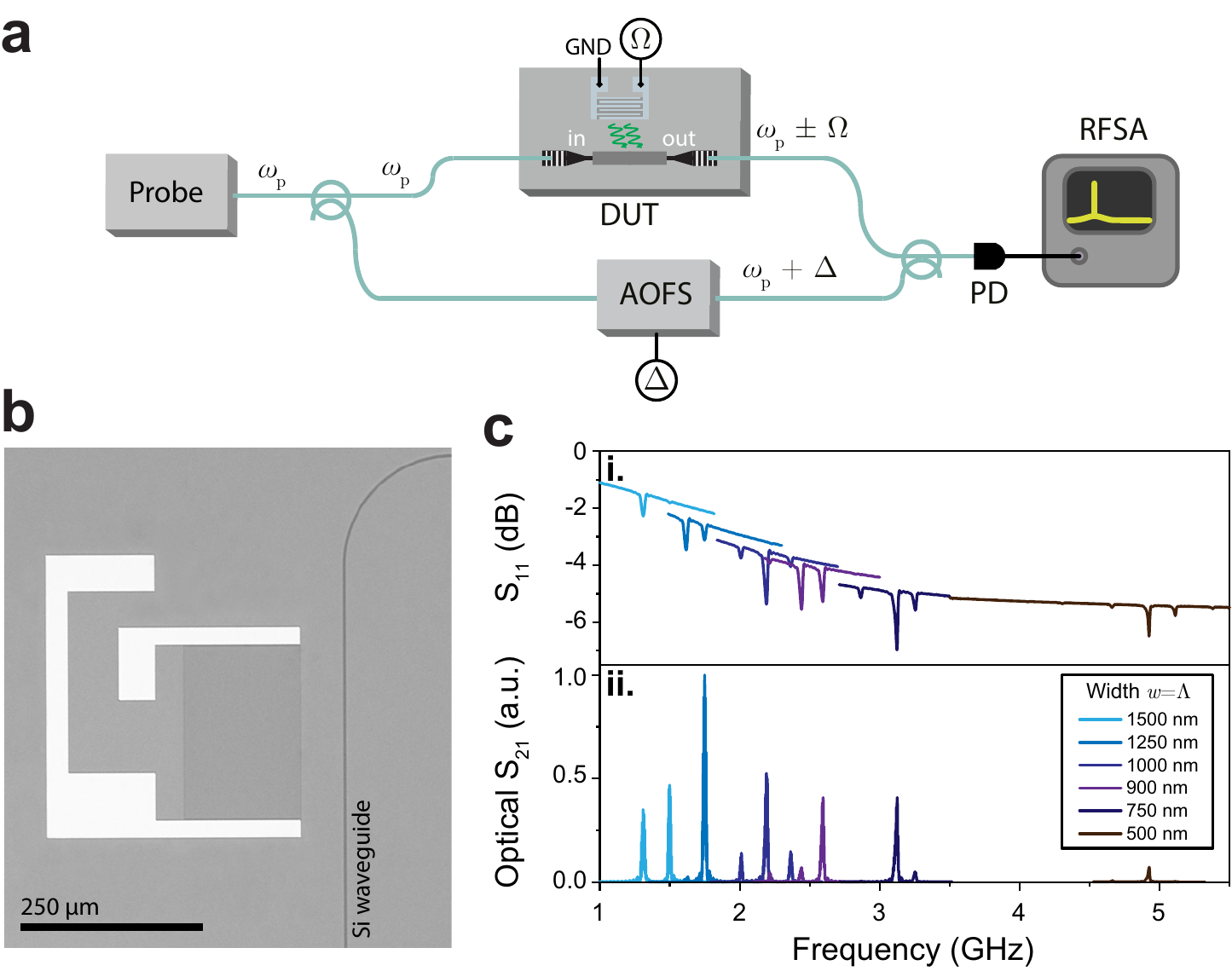}
\caption{Experimental observation of acousto-optic modulation in silicon. (a) depicts the experimental setup used to characterize the integrated AOM devices. Probe light at frequency $\omega_\textup{p}$ is split into two paths. In the lower path, it is frequency shifted to $\omega_\textup{p}+\Delta$ using an acousto-optic frequency shifter (AOFS), while in the upper path, light is coupled into the device under test (DUT). An on-chip IDT is driven by an external RF source at frequency $\Omega$. When $\Omega$ is near the operation frequency for the transducer, energy is converted to a SAW through the piezoelectric effect, which in turn mediates acousto-optic modulation of guided light. This modulation produces energy transfer to optical sidebands at $\omega_\textup{p} + \Omega$ and/or $\omega_\textup{p} - \Omega.$ Light is then coupled off chip, and combined with the frequency-shifted reference on a fast photodiode (PD). The resulting microwave spectrum is detected using a RF spectrum analyzer (RFSA). (b) Greyscale micrograph of a typical device, consisting of an IDT with contact pads in a ground-signal-ground configuration, and an integrated optical waveguide to the right of the transducer. (c.i) Electrical reflection coefficient $S_{11}$ as a function of drive frequency $\Omega$ for several devices with different values of IDT finger pitch $\Lambda$ and waveguide width $w$. (c.ii) Corresponding acousto-optic modulation coefficient $S_{21}$ for the same devices, demonstrating acousto-optic phase modulation in the range of $\Omega/2\pi = 1-5$ GHz.}
\label{fig:aomexp}
\end{figure*}

This AlN-on-SOI material stack allows separate fabrication of low-loss optical waveguides and efficient electromechanical transducers on the same platform. The silicon ridge waveguides guide TE-like optical modes with symmetric (Fig. \ref{fig:intro}d) and anti-symmetric (Fig. \ref{fig:intro}e) mode profiles through total internal reflection. At the same time, IDTs fabricated on top of the piezoelectric AlN layer enable electromechanical transduction of surface acoustic waves. The simulated elastic displacement profile of one such surface wave at frequency $\Omega/2\pi = 3.1$ GHz, which is preferably excited by a transducer with pitch $\Lambda = 750$ nm, is plotted in Fig. \ref{fig:intro}b-c. Importantly, the electromechanically-driven SAWs extend into the optical waveguide layer, and can modulate the effective indices of waveguide modes through the elasto-optic effect. The magnitude of the index modulation effect is $\propto n^3$ \protect{\cite{royer1999elastic}}, making silicon ($n = 3.48$) an excellent candidate for strong light-sound interactions, as has been previously demonstrated in nonlinear light-sound coupling through stimulated Brillouin scattering \protect{\cite{shinnatcomm,vanlaernatphoton,Kittlaus2016}}. In particular, the elastic wave in Fig. \ref{fig:intro}b-c has strong displacement that is co-polarized with the optical modes, allowing access to silicon's exceptionally strong $p_{11}$ photoelastic tensor component (for further details see Supplementary Note II).

The basic operation scheme of the acousto-optic modulator devices is depicted in Fig. \ref{fig:intro}f. Here, an electrical signal at frequency $\Omega$ drives a SAW at the same frequency using an IDT. This elastic wave is then incident on a silicon waveguide containing a light wave at frequency $\omega_\textup{p}.$ Elasto-optic modulation of the waveguide mode results in energy transfer to optical sidebands at frequencies $\omega_\textup{p} \pm \Omega$ through a phase modulation process. Later, we will also investigate acousto-optic single-sideband amplitude modulation through inter-modal light scattering, and study non-reciprocal acousto-optic modulators based on this process.

\begin{figure*}[htbp]
\centering%\vspace{-10pt}
\includegraphics[width=\linewidth]{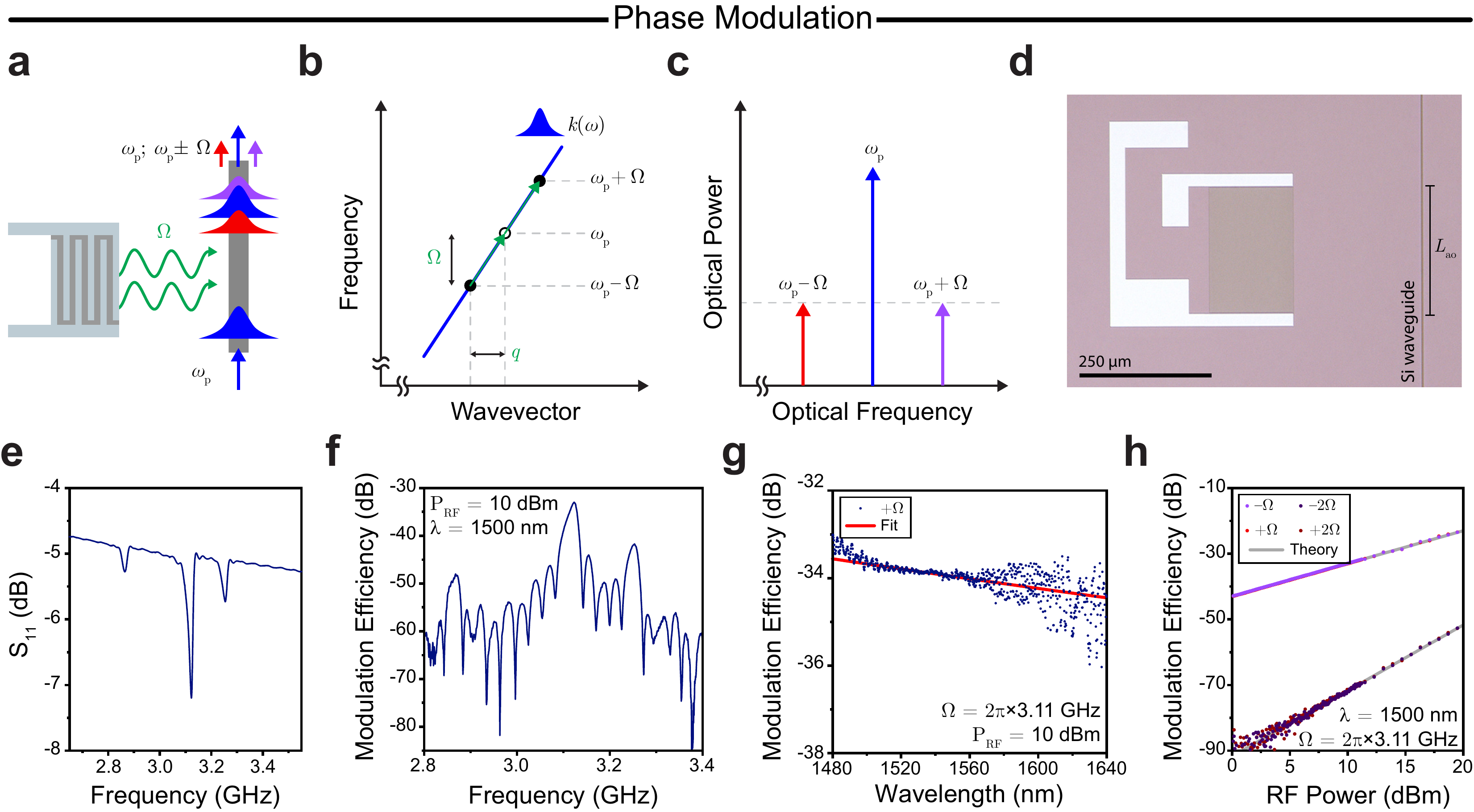}
\caption{Dynamics and experimental characterization of a representative acousto-optic phase-modulator device. (a) In a phase-modulation process, laterally-incident acoustic waves produce energy transfer from light at $\omega_\textup{p}$ to equal-amplitude sidebands at $\omega_\textup{p}\pm\Omega$ guided in the same optical mode. (b) plots the phase-matching diagram for this process, where an incident elastic wave with frequency $\Omega$ and axial wavevector $q$ scatters light up and down a single optical dispersion branch. Note that, because the incident SAW propagates normal to the direction of optical propagation, $q\approx0.$ (c) sketches the resulting phase-modulated optical spectrum as a function of optical frequency. (d) shows an optical micrograph of a fabricated phase-modulator device, with acoustic wavelength $\Lambda_\textup{ac} = 2\Lambda = 1.5$ $\upmu$m, waveguide width $w = 750$ nm, and acousto-optic interaction length $L_\textup{ao} \approx 0.24$ mm. (e) and (f) plot the electrical reflection coefficient for the IDT and acousto-optic scattering (modulation) efficiency (relative to the carrier power) as a function of drive frequency. In (g), the drive frequency is fixed to $\Omega/2\pi = 3.11$ GHz, and the modulation efficiency for the $+\Omega$ sideband as a function of optical wavelength is measured. Because the wavevector $q$ involved in the phase matching process is vanishingly small, modulation is optically broadband--a modest trend $\propto1/\lambda^2$, intrinsic to acousto-optic scattering processes, is observed, plotted as the red "fit" line. (h) plots the modulation efficiency as a function of RF drive power, showing a linear trend for the $\pm\Omega$ sidebands and quadratic trend for cascaded $\pm2\Omega$ sidebands, in agreement with phase modulator theory.} 
\label{fig:datapm}
\end{figure*}

\subsection{Observation of Acousto-optic Modulation}

Modulation of incident optical waves through the integrated AOM devices is studied using the experimental setup diagrammed in Fig. \ref{fig:aomexp}a. Probe light from a tunable semiconductor laser (frequency $\omega_\textup{p}$) is split into two paths; in the lower path, light is passed through an acousto-optic frequency shifter (AOFS) which shifts the optical frequency to $\omega_\textup{p} + \Delta$ (where $\Delta  = 2\pi\times100$ MHz) to serve as a frequency-detuned local oscillator for heterodyne detection. Meanwhile, in the upper path, light is coupled to the device using integrated grating couplers, where it is modulated by electrically-driven acoustic waves at frequency $\Omega$, transferring energy to optical sidebands at $\omega_\textup{p} + \Omega$ and/or $\omega_\textup{p} - \Omega$. After being coupled off-chip, this signal is combined with the local oscillator and incident on a fast photodiode, resulting in unique radiofrequency (RF) beat-notes at $\Omega\mp\Delta$ for the $\pm\Omega$-shifted sidebands, and $\Delta$ for un-shifted light, with the RF power in each line proportional to the optical power in the corresponding tone. 

We first examine on-chip acousto-optic interactions through elementary device designs of the type shown in the micrograph of Fig. \ref{fig:aomexp}b. Here, an IDT with pitch $\Lambda$ is used to excite surface acoustic waves with wavelength $\Lambda_\textup{ac} = 2\Lambda$. These SAWs are incident on an optical waveguide (with width $w=\Lambda=\Lambda_\textup{ac}/2$) in the transverse direction. As discussed in the next section, this geometry leads to acousto-optic phase modulation of guided optical waves, producing optical modulation sidebands of equal amplitude at $\omega_\textup{p} \pm \Omega$, where $\Omega$ is the drive frequency of the RF source.

%$\Omega/2\pi =$ 1$-$5.5 GHz

The electrical reflection coefficient $S_{11}$ of several IDTs, measured using an electrical network analyzer, is plotted in Fig \ref{fig:aomexp}c.i. as a function of drive frequency $\Omega/2\pi$. This plot reveals efficient transduction of acoustic waves as narrowband dips in the reflected power spectra when the drive frequency matches the frequency of propagating SAWs with acoustic wavelengths equal to the transducer pitch. (For more details on the IDT performance and design, see Supplementary Note I.) The corresponding modulation efficiency, in terms of the normalized optical sideband power (shorthand $S_{21}$), is plotted in Fig. \ref{fig:aomexp}c.ii. Acousto-optic modulation is observed at discrete peaks across this frequency window, corresponding to SAWs excited by the various IDTs; however the modulation efficiency drops above $\Omega/2\pi = 4$ GHz due to a combination of increased acoustic attenuation and reduced acousto-optic overlap; improved efficiency at higher frequencies should be possible by moving to thinner silicon dioxide and AlN films (for more details see Supplementary Note II).

%In the NIBS process, the travelling elastic wave breaks the symmetry between forward- and backward-propagating optical waves to produce unidirectional mode conversion and single-sideband modulation. This process is diagrammed in Fig. \ref{fig:intro}j; when light propagates in the forward direction within the modulator waveguide, it is mode-converted and frequency-shifted through a linear acousto-optic modulation process. However, when light is injected into the same waveguide in the backward direction, it propagates through the waveguide unaffected by the elastic wave (Fig. \ref{fig:intro}k).

\subsection{Intra-modal Optical Phase Modulation}

Thus far, we have described the AlN-on-SOI platform used to realize acousto-optic devices in silicon, and demonstrated acousto-optic modulation from 1-5 GHz frequencies. Next, we explore these interactions in detail, and show how the device design can be tailored to realize either intra-modal phase modulation (coupling light waves guided in a single spatial mode), or inter-modal single-sideband modulation. Thereafter, we will build on these fundamental capabilities to create and characterize a non-reciprocal modulator structure which produces both broadband, uni-directional light scattering, and enhanced acousto-optic coupling.

We first begin by studying the behavior of an acousto-optic phase modulator in Fig. \ref{fig:datapm}a-h. Through operation of this device, as shown in Fig. \ref{fig:datapm}a, elastic waves with frequency $\Omega$ are incident normal to the optical propagation axis (axial acoustic wavevector $q = 0$). Through elasto-optic refractive index modulation, this incident elastic wave scatters light guided in the symmetric optical waveguide mode at frequency $\omega_\textup{p}$ to frequency-detuned waves guided in the same optical mode at $\omega_\textup{p} \pm \Omega$.

Phase-matching for an acousto-optic scattering process requires that the sums of the wavevectors of the initial and final particle states are equal. For the $+\Omega$ (anti-Stokes, or blue-shifting) process, this requires that
\begin{equation}
k(\omega_\textup{p}+\Omega)  = k(\omega_\textup{p}) + q,
\label{eq1}
\end{equation}
whereas phase-matching for the $-\Omega$ (Stokes, or red-shifting) process requires that
\begin{equation}
k(\omega_\textup{p}-\Omega) + q = k(\omega_\textup{p}),
\label{eq2}
\end{equation}
where $k(\omega)$ is the wavevector of the symmetric optical mode at frequency $\omega$. These phase-matching conditions are plotted in frequency-wavevector coordinates in Fig. \ref{fig:datapm}b, showing optical scattering up and down the dispersion curve of the fundamental optical mode mediated by the incident elastic wave (green arrow). 

Provided that the optical group velocity $v_\textup{g}$ does not change appreciably over the frequency shift $\Omega$ (in other words that the dispersion is linear over frequency detunings of a few GHz), a condition well-satisfied in most waveguide systems, both Eqs. \ref{eq1}-\ref{eq2} can be approximated as $\Omega \partial k/\partial \omega = \Omega/v_\textup{g} = q$. Because the IDT in this system excites acoustic waves with identically zero wavevector ($q = 0$), the scattering process is not precisely phase-matched, with a wavevector mismatch $\Delta q$ given by      
\begin{equation}
\Delta q = \Omega/v_\textup{g} - q = \Omega/v_\textup{g} \approx 100 \;\textup{m}^{-1}.    
\end{equation}
As long as $\Delta q L/2 \ll 1$ is satisfied, where $L$ is the total device length, this phase mismatch is not appreciable over the interaction length, and the modulation process is approximately phase-matched. This is the case for our test structures, where L $\approx$ 240 $\upmu$m, so $\Delta q L/2 \approx .02$.  

Because phase-matching conditions for both Stokes and anti-Stokes processes are simultaneously satisfied in this device geometry, acousto-optic scattering to both sidebands occurs with equal efficiency, as diagrammed in Fig. \ref{fig:datapm}c. A closer examination reveals that these dynamics produce pure phase modulation \protect{\cite{royer1999elastic,kharelpra,GertlerNL2019}}. This can be understood from the fact that only the phase of the propagating optical modes is appreciably modulated through the elasto-optic interaction. %\hl{(is this true??)}.

Detailed experimental results for one device, with IDT pitch and acoustic wavelength $\Lambda = \Lambda_\textup{ac}/2 = 750$ nm  and optical waveguide width $w = 750$ nm are depicted in Fig. \ref{fig:datapm}e-h; an optical micrograph of this device is depicted in Fig. \ref{fig:datapm}d. This particular device has an approximate acousto-optic interaction length $L \approx 240$ $\upmu$m set by the transducer aperture, and number of IDT finger pairs $N = 107$. The electrical reflection coefficient $S_{11}$ of the IDT is plotted as a function of frequency in Fig \ref{fig:datapm}e, while the acousto-optic modulation efficiency is plotted in Fig. \ref{fig:datapm}f for an incident RF drive power of $P_\textup{RF} = 10$ mW. Note that, throughout this article, we define modulation efficiency as the ratio of the scattered optical sideband power relative to the incident probe power, i.e. efficiency $\eta^2 \equiv P_\textup{out}(\omega_\textup{p}-\Omega)/P_\textup{inc}(\omega_\textup{p})$. Fig. \ref{fig:datapm}f reveals acousto-optic modulation at frequencies corresponding to the dips in Fig. \ref{fig:datapm}e that represent efficient electro-mechanical transduction, including a strong peak around a center frequency of $\Omega/2\pi = 3.11$ GHz. The FWHM bandwidth $\Delta \Omega_\textup{ao}/2\pi = 20$ MHz and sinc-squared-like acousto-optic response result from the finite transducer length and equal finger weighting \protect{\cite{royer1999elastic}}. The approximate acoustic power launched toward the optical waveguide section is estimated from the dip depth $\Delta S_{11}$ of the electrical power reflection spectrum as $P_\textup{a} = 0.5  |\Delta S_{11}| P_\textup{RF} \approx 0.6$ mW. 

%Note that ripples in the acousto-optic response result from the fact that, in this device, the acoustic wave is actually incident in two separate waveguide sections (the latter of which is not shown in Fig. \ref{fig:datapm}e), forming an acoustic response reminiscent of a finite-impulse microwave photonic filter. 
Since the phase matching conditions in Eqs. \ref{eq1}-\ref{eq2} are approximately satisfied independent of optical wavelength, the phase modulation process is optically broadband; Fig. \ref{fig:datapm}g shows the measured modulation efficiency (blue dots) as a function of optical wavelength from $\lambda = 1480-1630$ nm, showing modest decrease $\propto \lambda^{-2}$ with increasing wavelength (red line) in good agreement with acousto-optic scattering theory \protect{\cite{royer1999elastic}}. 

The power dependence of the modulation efficiency is plotted in Fig. \ref{fig:datapm}h. At higher RF drive powers, cascaded sidebands appear as in a typical phase modulator response. The scattered optical power in the $\pm\Omega$ sidebands is linear with respect to RF drive power, confirming that the acousto-optic interaction is linear in nature. The strength of refractive index modulation in a modulator device, normalized to input voltage and device length, can be expressed in terms of a figure of merit $V_\pi L,$ where $V_\pi$ is the half-wave voltage and $L$ is the interaction length, equal to $L_\textup{ao}$ in these devices. The observed data correspond to $V_\pi L_\textup{ao} = 1.8$ V$\cdot$cm for this device, which compares well with results achieved in existing electro-optic modulator devices \protect{\cite{reed2010silicon, watts2010low, tbj2012, weigel2018bonded, wang2018integrated}}. Since the resulting light scattering rate is proportional to $L_\textup{ao}^2$, longer interaction lengths should dramatically improve the total modulation efficiency of future devices. 

\begin{figure*}[htbp]
\centering%\vspace{-10pt}
\includegraphics[width=\linewidth]{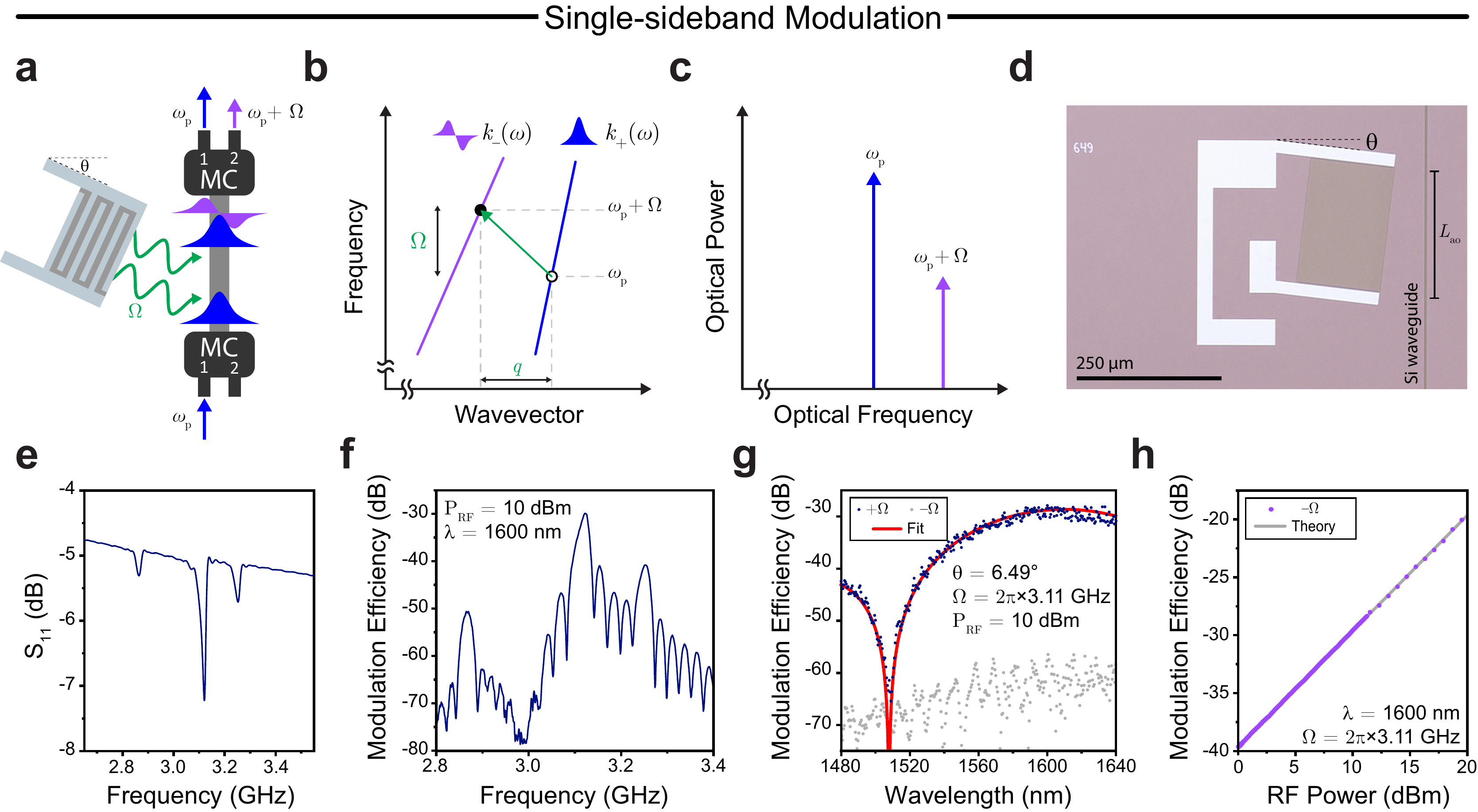}
\caption{Dynamics and experimental characterization of a representative acousto-optic single-sideband-modulator device. (a) Through a single-sideband modulation process, an elastic wave incident at angle $\theta$ (corresponding to nonzero axial wavevector $q$) and frequency $\Omega$ mode-converts and frequency-shifts a guided-wave optical signal. In the example diagram here, probe light at frequency $\omega_\textup{p}$ is injected into Port 1 of an integrated mode multiplexer (MC) which couples it into the symmetric waveguide mode. The incident acoustic wave scatters this optical wave to frequency $\omega_\textup{p}+\Omega$ in the anti-symmetric mode through a blue-shifting (anti-Stokes) process. This light then exits an identical mode multiplexer through Port 2, whereas unshifted probe light is de-multiplexed through Port 1. (b) plots phase-matching for this process, where the incident elastic wave mediates an inter-band transition between two distinct optical dispersion branches. In contrast to the phase-modulator (intra-modal) case, no additional scattering pathways exist, resulting in light scattering to a single frequency-detuned sideband, as diagrammed in (c); (d) shows an optical micrograph of a fabricated device, with IDT pitch $\Lambda = 747.5$ nm and acoustic wavelength $\Lambda_\textup{ac} = 1.5$ $\upmu$m, waveguide width $w = 1500$ nm, SAW angle $\theta=6.49^\circ$, and acousto-optic interaction length $L_\textup{ao} \approx 0.24$ mm. (e) and (f) plot the electrical reflection coefficient of the IDT and acousto-optic single-sideband modulation efficiency (relative to the carrier power) as a function of drive frequency for this device. In (g), the drive frequency is fixed to $\Omega/2\pi = 3.11$ GHz, and modulation efficiency for the $+\Omega$ tone is plotted as a function of wavelength. Because the inter-modal scattering process requires a specific elastic wavevector $q$, it is phase-matched around a specific optical frequency (wavelength), showing a sinc-squared response about a center wavelength of 1604 nm, as plotted in the red "fit" curve. The unwanted $-\Omega$ sideband amplitude is plotted in grey dots, showing strong single-sideband suppression of $\sim$30 dB. (h) plots the measured modulation efficiency as a function of RF power, showing a linear trend in sideband amplitude with increasing drive power.} 
\label{fig:dataam}
\end{figure*}

%experimental data typically show $\sim$30 dB relative suppression of the unwanted sideband, limited by modal crosstalk in the mode multiplexers.

\subsection{Inter-modal Single-sideband Modulation}

Next, we engineer inter-modal acousto-optic scattering in multimode optical waveguides. In contrast to the intra-modal (single-mode) process studied in the previous section, the inter-modal process scatters light waves between distinct waveguide modes, producing dynamics reminiscent of wavevector-selective Bragg scattering in bulk acousto-optic modulators. 

The inter-modal acousto-optic modulator is studied in Fig \ref{fig:dataam}a-h, with the basic device concept shown in Fig. \ref{fig:dataam}a. Here, the electrically-driven elastic wave propagates at an angle $\theta$ with respect to an optical waveguide, corresponding to a nonzero axial wavevector $q$. As a result of this nonzero $q$, the SAW can phase-match to inter-modal (or inter-band) photonic transitions where incident and modulated light are guided in distinct optical waveguide modes. 

To study inter-modal scattering, we interface a multimode optical waveguide with integrated mode multiplexers (labeled MC), as shown in Fig. \ref{fig:dataam}a. Light injected in Port 1 of a multiplexer is coupled into the symmetric optical mode, whereas light incident in Port 2 is coupled into the anti-symmetric mode. After these waves propagate through the waveguide, they can be de-multiplexed through an identical multiplexer into single-mode bus waveguides, and the optical waves are coupled off-chip. For details on mode multiplexer design, see Ref. \protect{\cite{Kittlaus2017}}.

A representative inter-modal transition is diagrammed in Fig. \ref{fig:dataam}a-b. As shown in Fig. \ref{fig:dataam}a, probe light at frequency $\omega_\textup{p}$ is injected into the symmetric waveguide mode of a multimode optical waveguide through an integrated mode multiplexer. An elastic wave with frequency $\Omega$ that is backward-propagating with respect to the optical waves ($q < 0$) is incident on this guided-wave optical signal. Through an inter-modal acousto-optic scattering process, probe light is mode-converted to the anti-symmetric waveguide mode and frequency-shifted to $\omega_\textup{p}+\Omega$. Afterwards, probe and modulated waves are spatially de-multiplexed through an identical mode multiplexer, which couples light remaining in the symmetric mode out of Port 1, and mode-converted light in the anti-symmetric mode out of Port 2.
%wavevector $k_+(\omega_\textup{p})$
%and wavevector $k_-(\omega_\textup{p}+\Omega)$ 

This operation scheme produces mode conversion and frequency-shifting of guided optical waves through the inter-band photonic transition diagrammed in Fig. \ref{fig:dataam}b, where the incident elastic wave (green arrow) with (frequency, wavevector) = $(\Omega, q)$ can be understood to scatter light between initial (open circle) and final (filled circle) optical states on two distinct optical dispersion branches. The phase-matching condition for this anti-Stokes scattering process reads
\begin{equation}
k_-(\omega_\textup{p}+\Omega) = k_+(\omega_\textup{p}) + q.
\label{ampm}
\end{equation}
where $k_+(\omega)$ and $k_-(\omega)$ refer to the dispersion relations for the symmetric and anti-symmetric optical modes, respectively. In contrast to the intra-modal modulation studied in Fig. \ref{fig:datapm}b, exactly two optical waves participate in this process, since the same elastic wave (green arrow in Fig. \ref{fig:dataam}b) cannot scatter light to any other available optical states through a phase-matched process \protect{\cite{Kittlaus2017}}. Therefore, this process produces single-sideband optical scattering, as plotted in Fig. \ref{fig:dataam}c. (Note that, were light originally injected into the opposite mode, a $-\Omega$ frequency shift would result, but the dynamics would remain otherwise identical. For further details, see Supplementary Note V.) This behavior is reminiscent of traditional bulk-crystal acousto-optic modulators, where Bragg scattering is used to favor single-sideband optical scattering to a specific diffraction angle (optical wavevector).  

As a result of optical dispersion, the inter-modal scattering process is phase-matched only around a particular probe frequency $\omega_0$ (wavelength $\lambda_0$) where Eq. \ref{ampm} is satisfied, determined through the transcendental equation 
%\frac{n_\textup{p,-}}{\lambda_0} - \frac{n_\textup{p,+}}{\lambda_0} = q,
%\begin{equation}
%\frac{\left(\omega_0 + \Omega \right)n_\textup{p,-}\left(\omega_0 +\Omega\right) - \omega_0 n_\textup{p,+}\left(\omega_0 %\right)}{c} = q
%\end{equation}
\begin{equation}
\frac{\omega_0 + \Omega}{c}n_\textup{p,-}\left(\omega_0 +\Omega\right) - \frac{\omega_0}{c} n_\textup{p,+}\left(\omega_0 \right) = q
\end{equation}
where $n_\textup{p,-}(\omega)$ and $n_\textup{p,+}(\omega)$ are the phase indices of the anti-symmetric and symmetric optical modes, and are functions of optical frequency $\omega$ (wavelength $\lambda$). According to this relationship, $\lambda_0$ is set through lithographic definition of the acoustic wavevector $q,$ which depends on the IDT angle $\theta$ and pitch $\Lambda$. For further details on the relationship between $q$ and design parameters, see Supplementary Note III.
%$\lambda_0 = \frac{2\pi}{q}\left(n^\textup{p}_-(\lambda_0) - n^\textup{p}_+  \right)$    

As the incident optical wavelength is detuned from $\lambda_0$, the scattering process accumulates a phase mismatch $\Delta q_\textup{pm}$ given by
\begin{equation}
\Delta q_\textup{pm} \approx \left(\frac{\partial k_-}{\partial \omega} - \frac{\partial k_+}{\partial \omega} \right)\Delta \omega = \frac{n_\textup{g,-}-n_\textup{g,+}}{c}\Delta \omega,    
\label{sincbw0}
\end{equation}
where $n_\textup{g,-}$ and $n_\textup{g,+}$ are the group indices of the anti-symmetric and symmetric optical modes. This phase mismatch results in a finite operation bandwidth of the inter-modal modulator:
\begin{equation}
\Delta \lambda_\textup{FWHM} = \frac{2\times1.39}{\pi \left | n_\textup{g,-}-n_\textup{g,+} \right |} \frac{\lambda_0^2}{L},
\label{sincbw}
\end{equation}
where the factor of 1.39 comes from the shape of the resulting sinc-squared response (for more details see Supplementary Note IV). For a device of interaction length $L = 240$ $\upmu$m, and $\left | n_\textup{g,-}-n_\textup{g,+} \right | = 0.12,$ corresponding to the designs of our devices, $\Delta \lambda_\textup{FWHM} = 75$ nm. 

Beyond requiring strict phase-matching, an additional condition for efficient inter-modal acousto-optic scattering is that the incident elastic wave should have the correct shape and symmetry to mediate efficient acousto-optic coupling between the two guided optical modes. For the symmetric and anti-symmetric modes guided in our silicon waveguides (Fig. \ref{fig:intro}d-e), simulations show that acousto-optic coupling is optimal when transverse acoustic wavelength $\Lambda_\textup{ac}$ is around the waveguide width $w$. The elastic wave plotted in Fig. \ref{fig:intro}b-c is one such example. In this case, the IDT pitch $\Lambda = \Lambda_\textup{ac} \cos(\theta)/2,$ and the axial wavevector $q = (2\pi/\Lambda_\textup{ac}) \tan(\theta) = (\pi/\Lambda) \sin(\theta).$ For further details on acousto-optic overlap and IDT design, see Supplementary Notes I-III and Refs. \protect{\cite{Tadesse2014,sohn17,Kittlaus2018}}.  

We fabricated single-sideband modulator devices of various widths (acoustic wavelengths) ranging from $w=\Lambda_\textup{ac}=1-1.5$ $\upmu$m on the AlN-SOI platform. Experimental results for one such device with IDT pitch $\Lambda = 747.5 $ nm ($\Lambda_\textup{ac} = 1500 $ nm) and optical waveguide width $w = 1500$ nm are depicted in Fig. \ref{fig:dataam}e-h; an optical micrograph of the interaction region is shown in Fig. \ref{fig:dataam}d. This device has an approximate interaction length $L \approx 240$ $\upmu$m set by the transducer aperture, number of transducer finger pairs N = 107, and IDT angle $\theta = 6.49^\circ.$ The electrical power reflection spectrum $S_{11}$ for the angled IDT is plotted in Fig \ref{fig:dataam}e, and is practically identical to that of straight transducer in Fig. \ref{fig:datapm}e. The acousto-optic modulation efficiency at $\lambda = 1600$ nm, in terms of scattered sideband power relative to the incident probe power, is plotted in Fig. \ref{fig:dataam}f for an incident RF drive power of 10 mW, revealing efficient acousto-optic modulation around a center frequency of $\Omega/2\pi = 3.11$ GHz. The FWHM bandwidth of the electrical modulation response $\Delta \Omega_\textup{ao}/2\pi = 20$ MHz. As before, the approximate acoustic power launched toward the optical waveguide section is $P_\textup{a} = 0.6$ mW. 

We study the wavelength-dependence of the acousto-optic response and single-sideband modulation in Fig. \ref{fig:dataam}g. As expected from Eqs. \ref{sincbw0}-\ref{sincbw}, modulation is observed for a broad bandwidth around a central wavelength of $\lambda_0 = 1604$ nm, set lithographically by the IDT angle $\theta = 6.49^\circ.$ The red fit curve in Fig. \ref{fig:dataam}g corresponds to a sinc$^2$ response with a FWHM modulation bandwidth $\Delta \lambda_\textup{FWHM} = 84$ nm, and the trend shows excellent agreement with the measured data. Efficient light scattering is observed only to the $+\Omega$ sideband (blue dots) through this single-sideband modulation process. In practice, a small amount of crosstalk in the mode multiplexers results in scattering to the unwanted $-\Omega$ optical sideband, plotted in light grey in Fig. \ref{fig:dataam}g. The measured single-sideband suppression for this device is around $-30$ dB or better across the operation wavelength range. %For details on the mode multiplexer design, see Supplementary Note R. %, as shown in the xx circles of Fig. xx. (should probably add a data scan)

%Fabricated devices are interfaced with mode-selective optical couplers (not shown; discussed in detail in references XX, YY), which are used and mode (de-)multiplexers which separately access the symmetric and anti-symmetric optical waves. (For details on multiplexer design and performance, see Supplementary Section Sx). 

The power-dependence of the scattered sideband amplitude is plotted in Fig. 3h, and is linear in RF drive power. In contrast to the phase modulator devices, no cascaded scattering is observed. At the highest tested power $P_\textup{RF} = 89$ mW (acoustic power $\approx$6 mW), a scattering efficiency of around 1\% (-20 dB) is observed for this $L = 240$ $\upmu$m-long device. 

\begin{figure*}[htbp]
\centering%\vspace{-10pt}
\includegraphics[width=\linewidth]{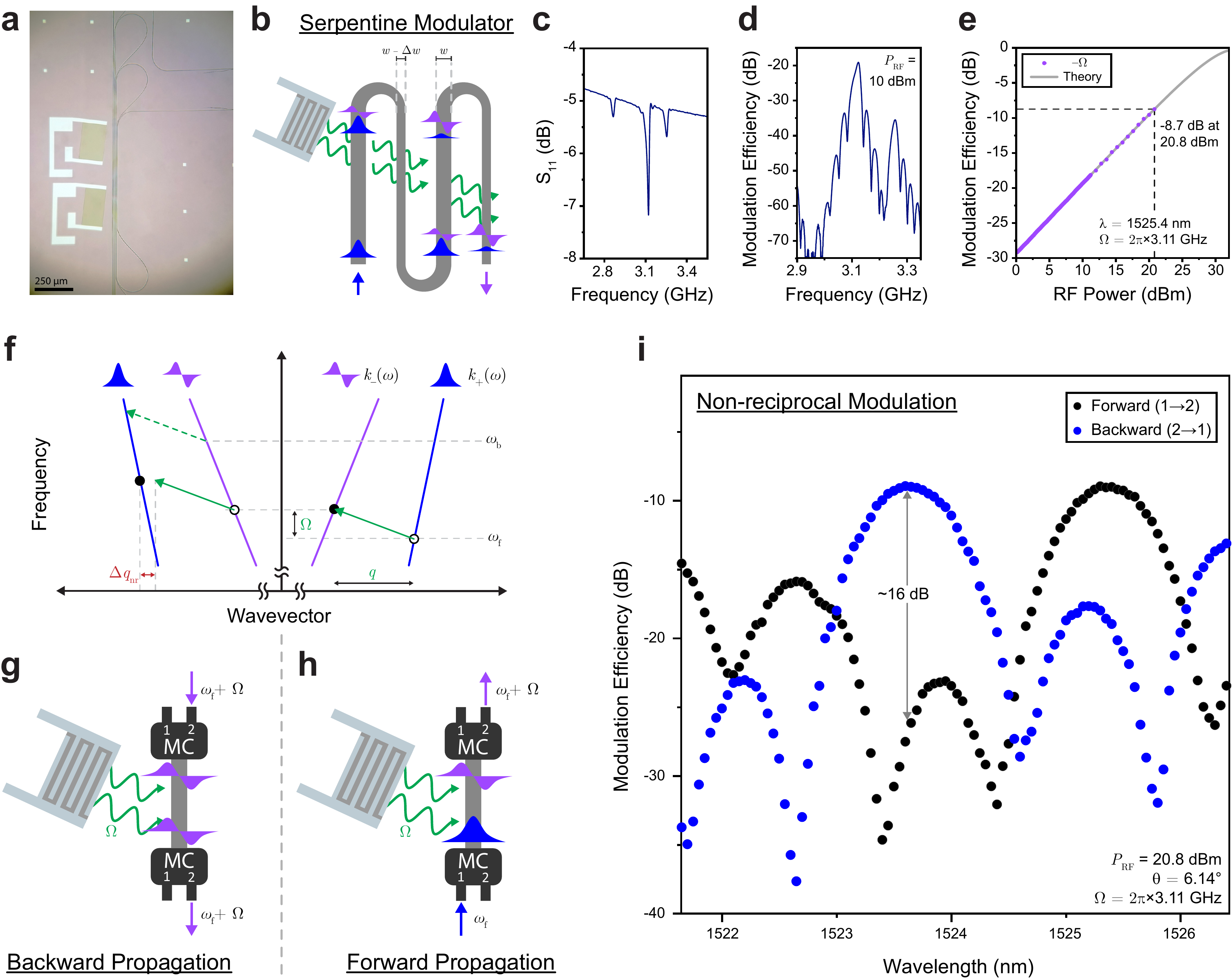}
\caption{Enhanced modulation efficiency and non-reciprocal mode conversion in a serpentine modulator device. (a) Micrograph of a fabricated serpentine modulator device consisting of a tightly-wrapped waveguide structure. Only the top IDT is used throughout these experiments. (b) Operation concept of the serpentine modulator device. When optical and acoustic waves are counter-propagating, the waveguide is designed to have width $w,$ and phase-matching is satisfied for efficient acousto-optic modulation. In the counter-propagating regions, the waveguide is tapered down to $w-\Delta w,$ changing the propagation constants of the optical modes so that phase matching is not satisfied, and hence modulation is inhibited. (c) plots the measured electrical reflection spectrum $S_{11}$ of the IDT, while (d) shows the acousto-optic response as a function of drive frequency $\Omega$. (e) plots the modulation efficiency as a function of RF drive power when $\Omega/2\pi = 3.11$ GHz and $\lambda = 1525.4$ nm, showing a $>$10-fold enhancement over the shorter devices. (f-h) diagram the origin of non-reciprocal modulation in these devices. In the forward direction, shown in (h) and the righthand side of (f), inter-modal acousto-optic scattering is phase-matched around optical frequency $\omega_\textup{f},$ where the incident elastic wave (green arrow) scatters light between inital (open circle) and final (filled circle) states on the optical dispersion bands. This leads to mode conversion and frequency shifting within the modulator device. In the backward direction, shown in (g) and the lefthand side of (f), a non-reciprocal wavevector mismatch $\Delta q_\textup{nr}$ arising from the nonzero group velocities of the optical modes prevents phase-matching, and no modulation occurs within the device. However, modulation may be phase-matched around a nearby frequency, $\omega_\textup{b}.$ Note that in (f-g), we considered back-scattered light at a frequency $\omega_\textup{f}+\Omega,$ since this the frequency of output light in (h) that could be back-scattered into the device by an external scatterer. However, this device supports broadband operation such that the response is similar for back-injected light in a wide band around $\omega_\textup{f}.$ (i) shows the experimental observation of non-reciprocal modulation, showing significant non-reciprocal contrast across the device's 0.8 nm (100 GHz) operation bandwidth.}
\label{fig:nonre}
\end{figure*}

\subsection{Non-reciprocal Modulation and Enhanced Efficiency}

We have demonstrated acousto-optic modulation using the AlN-on-SOI platform and shown how guided-wave acousto-optic interactions can be engineered to produce optical phase modulation and single-sideband modulation. In this section, we show how these couplings can be extended to longer device lengths, dramatically enhancing net modulation strength, and enabling non-reciprocal modulation and mode conversion. This traveling-wave non-reciprocal modulator represents a fundamental building block for silicon-based acousto-optic isolators and circulators.

We demonstrate non-reciprocal modulation using a serpentine single-sideband modulator which permits acousto-optic coupling over long interaction lengths. One such device is shown in the micrograph of Fig. \ref{fig:nonre}a. The basic design concept of this serpentine single-sideband modulator is depicted in Fig. \ref{fig:nonre}b; an inter-digitated transducer emits elastic waves which traverse a serpentine waveguide structure. When injected light waves are counter-propagating with respect to the SAW, phase-matching is satisfied for inter-modal light scattering, and the optical waveguide has lithographically-defined width $w$. However, when light waves are co-propagating with the SAW, (i.e. after a waveguide bend), the waveguide is tapered to a narrower width $w-\Delta w,$ changing the effective indices of the propagating optical modes so that (1) phase-matching is not satisfied, and no appreciable acousto-optic interaction occurs, and (2) evanescent coupling between adjacent waveguide sections is inhibited. This geometry preserves the single-sideband nature of the inter-modal scattering process, but allows several successive lengths of optical waveguide to interact with acoustic waves generated in the same IDT.

A fabricated device is shown in Fig. \ref{fig:nonre}a. This device, with waveguide width $w = 1.5$ $\upmu$m and taper asymmetry $\Delta w = 35$ nm, is designed so that four separate waveguide sections interact with the same acoustic wave, resulting in an effective coupling length $L_\textup{ao} = 0.96 $ mm. The total device length from the beginning to end of acousto-optic coupling is $L_\textup{tot} = 1.53 $ cm. As shown in Fig. \ref{fig:nonre}a, the same device can be interfaced with multiple acoustic transducers to produce different modulation responses. For these measurements, we use the upper IDT with angle $\theta = 6.14^\circ,$ corresponding to a center operating wavelength of $\lambda_0 = 1525.4$ nm. The electrical power reflection spectrum of this device in Fig. \ref{fig:nonre}c is similar to that of prior devices. 

We first examine the modulation efficiency of this device at a wavelength of $\lambda_0 = 1525.4$ nm. The frequency-dependent modulation efficiency at an RF drive power of 10 mW (acoustic power $\approx$0.6 mW) is plotted in Fig. \ref{fig:nonre}d, showing a $\sim$11 dB increase in single-sideband scattered power over the shorter device in Fig. \ref{fig:dataam}f. This agrees reasonably well with the expected increase $\propto L_\textup{ao}^2$ over the shorter device. Next, the drive frequency is fixed to $\Omega/2\pi = 3.11$ GHz, and the RF drive power is increased, as shown in Fig. \ref{fig:nonre}e. At the highest tested power of $P_\textup{RF} =$ 20.8 dBm (120 mW), a single-sideband modulation efficiency $P(\omega_\textup{p}+\Omega)/P(\omega_\textup{p}) = -8.7 $ dB (13.5\%) is achieved.   

The increased propagation length $L_\textup{tot}$ of this serpentine modulator design enables a powerful form of traveling-wave non-reciprocal modulation, previously studied in Refs. \protect{\cite{Yu2009,lira12,Kittlaus2018}}. Through this interaction, the inter-modal scattering process is phase-matched only in a single propagation direction, enabling unidirectional modulation as a basis for optical isolators and circulators. 

%%
%%%%%%%%%%%%%%%%%%%%%%%
%This type of non-reciprocal modulation typically requires interactions lengths of beyond 1 cm \protect{\cite{lira12,Kittlaus2018}}, or the use of optical ring resonators to achieve sufficient wavevector resolution \protect{\cite{sohn2017}}. However, here we demonstrate that the same physics are possible with shorter modulation lengths, provided that there is a sufficient total optical propagation length between modulation stages.

The origin of non-reciprocal modulation can be understood through the phase matching diagram in Fig. \ref{fig:nonre}f, which plots the optical dispersion relations for light propagating in both the forward (righthand side) and backward (lefthand) directions. For simplicity, we consider the anti-Stokes process, though identical dynamics are produced for a Stokes (red-shifting) process. For light injected in the forward direction into the symmetric mode, as shown in Fig. \ref{fig:nonre}h, the inter-modal modulation process is phase-matched around an optical frequency $\omega_\textup{f},$ resulting in mode conversion into the anti-symmetric mode and frequency-shifting to $\omega_\textup{f}+\Omega.$ However, for light injected backward into the anti-symmetric mode around the same frequency, the nonzero group velocities (dispersion) of the propagating modes result in a non-reciprocal wavevector mismatch $\Delta q_\textup{nr}$, given by \protect{\cite{Kittlaus2018}}:

\begin{equation}
\Delta q_\textup{nr} \approx \frac{\Omega}{c}\left(n_\textup{g,-} + n_\textup{g,+} \right).
\end{equation}

Provided that $\Delta q_\textup{nr} L_\textup{tot} \gg 1$, this process is not phase-matched around $\omega_\textup{f}$. In this case, light injected in the backward direction passes through the device without experiencing modulation, shown in Fig. \ref{fig:nonre}g. Interestingly, due to differing dispersion curve slopes (optical group velocities) between the two optical modes, the scattering process may be phase-matched for backward-propagating light at a nearby optical frequency $\omega_\textup{b}$, as shown by the dashed green arrow in Fig. \ref{fig:nonre}f. The frequency-splitting between forward- and backward-modulation frequencies can be calculated by linearizing the optical dispersion:
\begin{equation}
\omega_\textup{f}-\omega_\textup{b} = \frac{n_\textup{g,-} + n_\textup{g,+}}{n_\textup{g,-} - n_\textup{g,+}}\Omega.    
\end{equation}
Provided that $\omega_\textup{f}-\omega_\textup{b}$ is much greater than the FWHM operation bandwidth of the modulator device, strong non-reciprocity should be supported. Interestingly, this condition turns out to be functionally identical to the condition that the non-reciprocal phase mismatch $\Delta q_\textup{nr} L_\textup{tot}$ is large. Through the use of a serpentine design, long total interaction lengths $L_\textup{tot}$ are possible even when the acousto-optic coupling occurs only over relatively short waveguide segments (see Supplementary Note IV for details). For our device, $\Delta q_\textup{nr} L_\textup{tot} \approx 9$ indicating that modulation should be non-reciprocal.

To experimentally characterize the non-reciprocal response of our devices, we fix the drive frequency to $\Omega/2\pi = 3.11$ GHz and inject probe light in either the forward and backward directions of a single device. The resulting modulation efficiency as a function of optical wavelength is plotted in Fig. \ref{fig:nonre}i, demonstrating non-reciprocal transmission around both $\lambda_\textup{f} = 2\pi c/\omega_\textup{f} = 1525.4$ nm and $\lambda_\textup{b} = 2\pi c/\omega_\textup{b} = 1523.7$ nm. Notably, this electrically-driven non-reciprocal modulator provides about 15 dB of non-reciprocal contrast between forward- and backward-propagating waves over the entire FWHM bandwidth of 0.8 nm (100 GHz) around $\lambda_\textup{b} = 1523.7$ nm. Both the non-reciprocal contrast and total modulation efficiency can be improved by moving to longer device lengths, while the operating bandwidth can be increased through optical dispersion engineering \protect{\cite{Yu2009,Kittlaus2018}}. 

%The origin of the nonreciprocal modulation response in this system can be understood from the distinct phase-matching requirements for inter-band scattering in the forward and backward directions. In this section, we explore the response of this system and determine the necessary conditions for wide-band nonreciprocal operation.

\section{Discussion}
We fabricated electromechanical SAW transducers on an AlN-on-SOI material stack, and used this platform to demonstrate acousto-optic modulation in integrated silicon waveguides. Harnessing intra- and inter-modal scattering processes, we separately engineered the acousto-optic interaction to produce optical phase modulation or wavevector-selective single-sideband amplitude modulation with $\geq$ 30-dB of single-sideband suppression. The latter represents a practical analogue of bulk acousto-optic frequency shifters implemented within integrated photonics, opening the door to a variety of applications including signal (de-)modulation, sensing, waveform synthesis, and on-chip heterodyne detection. By scaling the acousto-optic interaction to longer lengths in a serpentine modulator structure, we demonstrated efficient light scattering and electrically-driven non-reciprocal modulation in silicon. These compact AOM devices require no optical pumping of acoustic waves and no suspended structures, and utilize elasto-optic modulation within standard silicon waveguides.

These first-generation modulators exhibited reasonable modulation efficiencies of $1-13.5\%$ over short ($<$1 mm) acousto-optic coupling lengths, which can be immediately improved through several strategies. First, due to the ultra-low propagation losses of the silicon ridge waveguides, these interactions can easily be extended to longer interaction lengths (measured losses of the symmetric mode: $<$0.15 dB/cm; anti-symmetric mode: $<$0.35 dB/cm at waveguide width $w = 1.5$ $\upmu$m). For example, extended serpentine structures can be designed to achieve acousto-optic modulation (with efficiency $\propto L_\textup{ao}^2$) over cm-scales with negligible optical insertion losses. Second, power efficiency may be improved through the use of improved IDT designs over the simple inter-digitated electrodes used here. For example, structural and impedance-matching optimizations, the use of unidirectional transducers, or the design of SAW transducers which couple more efficiently to elastic modes with strong acousto-optic overlap could all improve modulation efficiency dramatically. (For more details on acousto-optic overlap calculations, see Supplementary Note II.) Finally, modulation strength may also be improved through the use of SAW cavities, edge reflectors or other forms of acoustic reflectors or resonators, or through the use of materials with stronger piezo-electric coupling such as ScAlN \protect{\cite{hashimoto2013high}}. Based on these types of incremental improvements, modulator devices that permit near-unity modulation efficiency at mW-level RF powers should be within reach.

Building on these foundational results, we developed distributed, non-reciprocal acousto-optic modulators, enabling uni-directional modulation and mode conversion with a combination of significant non-reciprocal contrast ($\sim$15 dB), low insertion losses ($<$0.6 dB), broad optical bandwidth (100 GHz; 0.8 nm), and high modulation efficiency ($>10^{-1}$), all in a silicon waveguide platform. These results represent a significant advance toward the creation of practical, non-magnetic integrated isolators and circulators \protect{\cite{Yu2009,lira12,sohn17,Kittlaus2018}}. Under its current design, the serpentine single-sideband modulator already represents a frequency-shifting four-port circulator \protect{\cite{Kittlaus2018}}. As future device designs push modulation efficiency close to unity, this operation can be readily adapted to create an optical isolator by routing either modulated or un-modulated light into an optical absorber or drop port. Frequency-neutral operation in optomechanical circulator devices can alternately be implemented through several successive compensating frequency shifts, if necessary. In contrast to existing approaches for optomechanical non-reciprocity \protect{\cite{Shen2016,Ruesink2016,Kim2017}}, this traveling-wave, non-reciprocal AOM supports broadband operation and avoids the use of optical resonators or suspended mechanical structures. With the improvements in efficiency discussed above, this platform may enable low-loss, broadband acousto-optic isolators and circulators with sub-mW power consumption.

In summary, we have demonstrated strong electrically-driven acousto-optic modulation in integrated silicon waveguides. Using this approach, we realized acousto-optic phase modulation, single-sideband amplitude modulation, and broadband ($>$100 GHz) non-reciprocal light propagation. The implementation of electro-mechanical transducers in a piezoelectric top cladding can be extended to other material systems as a general approach for realizing linear acousto-optic interactions in integrated photonic circuits. Because the optical waveguides are spatially separated from the acoustic transducers, we preserve the ultra-low losses of buried silicon waveguides as a path to efficient and low-loss integrated AOMs. These results represent a promising technique to add miniaturized acousto-optic modulators, frequency-shifters, and non-reciprocal components to the toolkit of integrated photonics. 

\section{Methods}

\subsection{Device Fabrication}
The integrated acousto-optic modulators were fabricated through a multi-layer lithography process. First, ridge waveguides were patterned in hydrogen silesquioxane (HSQ) using electron beam lithography. After development in TMAH, a Cl$_2$/BCl$_3$ dry etch was used to etch the ridge waveguides and grating couplers with a 90 nm etch depth. The remaining HSQ was removed in 5\% HF, and a 700 nm silicon dioxide overcladding was deposited using plasma-enhanced chemical vapor deposition. Following an anneal at 750$^\circ$ for four hours to improve the silicon dioxide purity, 600 nm of piezoelectric c-axis-oriented AlN was deposited using an AC magnetron sputtering process (OEM Group). Finally, inter-digitated transducers were patterened in 50 nm evaporated Al metal using a liftoff process on ZEP520A photoresist.

\subsection{Experiment}
The IDTs are excited using RF signals from an analog signal generator (Keysight E8257D), and heterodyne measurements of acousto-optic modulation are analyzed using a RF signal analyzer (Keysight N9030B). To measure the electrical power reflection $S_{11}$ of the IDT devices, the RF synthesizer is replaced with a calibrated electrical vector network analyzer (Keysight E5071C). For modulation efficiency measurements with incident RF power $>$10 dBm, an additional low-noise microwave amplifier is used (Mini-Circuits ZX60-83LN-S+). Optical interrogation of the AOM devices is achieved using a semiconductor laser (Santec TSL-710) that is tunable between 1480-1640 nm. Light is coupled on- and off-chip through integrated grating couplers interfaced with commercial four-port fiber arrays (OZ Optics). Typical fiber-to-chip coupling losses are $<$6 dB/facet at 1535 nm. 

\subsection*{Acknowledgements}
This research was carried out at the Jet Propulsion Laboratory, California Institute of Technology, under a contract with the National Aeronautics and Space Administration. The authors acknowledge helpful discussions with S. Gertler, P.O. Weigel, M.S. Mohamed, S. Forouhar, L. Sterczewski, A. Qamar, C. Frez, and D. Wilson. N.T.O. acknowledges support from the National Science Foundation Graduate Research Fellowship under grant no. DGE1122492.

\subsection*{Author Contributions}
E.A.K. and P.T.R. conceived the project and developed numerical and analytical models of the device physics. E.A.K. designed the devices with the assistance of P.T.R, N.T.O., R.E.M., and M.R.. E.A.K., W.M.J., and R.E.M. fabricated the devices. E.A.K. conducted the experiments with the assistance of W.M.J., N.T.O., and M.R. All authors contributed to the writing of this paper.

%\subsubsection*{Additional Information}
%\noindent \textbf{Copyright} \textcopyright \thinspace 2020. All rights reserved.

%\noindent \textbf{Competing financial interests:} The authors declare no competing financial interests.

%\noindent \textbf{Data availability:} All data supporting the findings of this work are available within the article and its supplementary information files.

%\bibliographystyle{naturemag-ed} 
%\bibliography{AOMPaper}

\newpage
\clearpage



\section*{Supplementary material}

# Supplementary Information: Electrically-driven Acousto-optics and Broadband Non-reciprocity in Silicon Photonics

Eric A. Kittlaus,[1] William M. Jones,[1] Peter T. Rakich,[2] Nils T. Otterstrom,[2] Richard E. Muller,[1] and Mina Rais-Zadeh[1]

[1]*Jet Propulsion Laboratory, California Institute of Technology, Pasadena, CA 91109 USA.*

[2]*Department of Applied Physics, Yale University, New Haven, CT 06520 USA.*

## Contents

# I. INTER-DIGITATED TRANSDUCER MODEL AND DESIGN

To facilitate efficient electro-mechanical transduction of surface acoustic waves (SAWs), inter-digitated transducers (IDTs) are fabricated in a 50-nm-thick Al metal layer on top of the AlN-SOI material platform. Representative micrographs at various levels of detail are shown in Fig. 1a-c, demonstrating good dimensional control and fabrication quality. In the scanning electron micrograph (SEM) of Fig. 1c, both the individual crystals of AlN and Al metal grains are visible.

The electro-mechanical response of our inter-digitated transducers (IDTs) is well-described by a Mason equivalent circuit model [1, 2], where the IDT response consists of a static capacitance $C_\mathrm{T}$ from the inter-digitated finger array, and a radiation impedance resulting from coupling to SAWs. The total electrical admittance of the transducer is given by

$$Y_\mathrm{T} = 2\pi i f C_\mathrm{T} + G_\mathrm{a}(f) + i B_\mathrm{a}(f), \tag{1}$$

where the radiation conductance $G_\mathrm{a}(f)$ and motional susceptance $B_\mathrm{a}(f)$ comprise the electromechanical response, and are given by

$$G_\mathrm{a}(f) = 8 K^2 C_\mathrm{T} N f_0 \left(\frac{\sin X}{X}\right)^2 \equiv G_0 \left(\frac{\sin X}{X}\right)^2, \tag{2}$$

$$B_\mathrm{a}(f) = G_0 \left(\frac{\sin 2X - 2X}{2X^2}\right), \tag{3}$$

and

$$X \equiv N\pi \frac{|f - f_0|}{f_0} \tag{4}$$

is the frequency detuning from the center frequency $f_0$. In these equations, $K^2$ is the material electromechanical coupling coefficient, and $N$ is the number of transducer finger pairs. Furthermore, for an inter-digitated capacitor with 50% duty cycle (filling factor), $C_\mathrm{T}$ can be calculated as [1]:

$$C_\mathrm{T} = N w_\mathrm{T} \left(\varepsilon_0 + \varepsilon_\mathrm{p}\right), \tag{5}$$

where $w_\mathrm{T}$ is the transducer width (aperture) given by the finger overlap length, and $\varepsilon_\mathrm{p}$ is the effective permittivity of the substrate. For our transducers with $w_\mathrm{T} \approx 240$ μm and $\varepsilon_\mathrm{p} \approx 9\varepsilon_0$, we calculate a static capacitance per finger pair of $C_\mathrm{T}/N = .021$ pF/N, or around $C_\mathrm{T} = 2.25$ pF for a transducer with $N = 106$.

Our transducers exhibit a parasitic resistance $R_\mathrm{p}$ resulting from the finite conductivity of the aluminum metal, and a small parasitic inductance $L_\mathrm{p}$, which are captured by the series circuit model of Fig. 1d. Using a calibrated electrical vector network analyzer, the total impedance $Z$ of our IDTs is measured as a function of frequency, as plotted in Fig. 1e.i-ii. These wideband data are fit to determine $R_\mathrm{p}$ (real part of $Z$), and $L_\mathrm{p}$ and $C_\mathrm{T}$ (imaginary part of $Z$). For the representative transducer studied in Fig. 1, with pitch $\Lambda = 750$

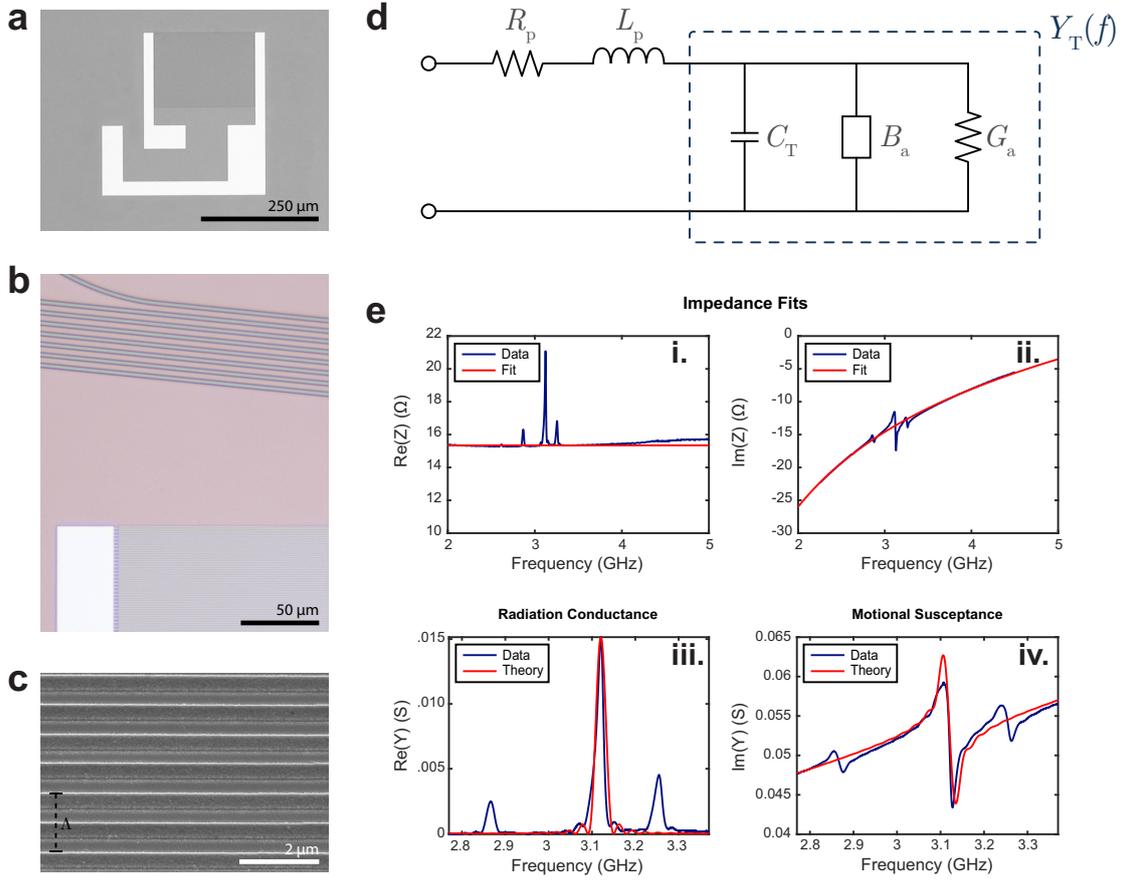

FIG. 1: Characterization of inter-digitated SAW transducers (a) depicts a greyscale micrograph of a representative IDT device fabricated in Al metal on AlN-SOI. The electrical contact pads are fabricated in the same layer in a ground-signal-ground (GSG) configuration. (b) shows a zoomed-in view of a transducer corner near a serpentine modulator device, resolving the individual 375-nm-wide fingers making up the $\Lambda = 750$ nm transducer. (c) shows a scanning electron micrograph of several IDT fingers on top of poly-crystalline AlN as viewed at a 30-degree angle with respect to normal. (d) diagrams an equivalent electrical circuit that describes the electrical behavior of the IDT. This circuit consists of external resistances and inductances $R_p$ and $L_p$, and a transducer with frequency-dependant admittance $Y_t(f)$. The transducer is specifically modeled as a parallel combination of a static capacitance $C_T$ from the inter-digitated fingers, and a radiation conductance $G_a(f)$ and motional susceptance $B_a(f)$ due to the electromechanical response. (e)i.-ii. plot the measured wideband impedance of the IDT as a function of frequency, which is used to determine $C_T$, $R_p$, and $L_p$ through fitting. The remaining panels plot $G_a(f)$ and $B_a(f)$ as calculated from the narrowband data and these fit parameters (dark blue) or according to a simple transducer model (red line).

nm, number of finger pairs $N = 106$, and aperture width $w_T \approx 240$ μm, we measure values of $R_p = 15.3$ Ω, $C_T = 2.72$ pF, and $L_T = 0.26$ nH.

Based on our experiments with different transducer designs and dimensions on this material platform, $L_T$ is consistently small, and decreases slightly with decreasing electrical contact area. $C_T$ is nearly linearly proportional to $N$, as predicted by Eq. 5, though usually slightly larger than predicted by theory, perhaps indicative of a larger value of $\varepsilon_p$.

The fit values of these circuit parameters are combined with the model of Eqs. 2-3 and narrowband

impedance measurements of $Y_T$ to determine $G_0$. For the device in Fig. 1e, $G_0 = 0.015$ S for the most strongly-coupled SAW mode, corresponding to an electromechanical coupling coefficient $K^2 = 0.22\%$ at $f_0 = 3.11$ GHz. This number is comparable to that observed in bulk AlN, indicating good film quality, and may be further enhanced due to the multi-layer acoustic structure and presence of the silicon device layer [3].

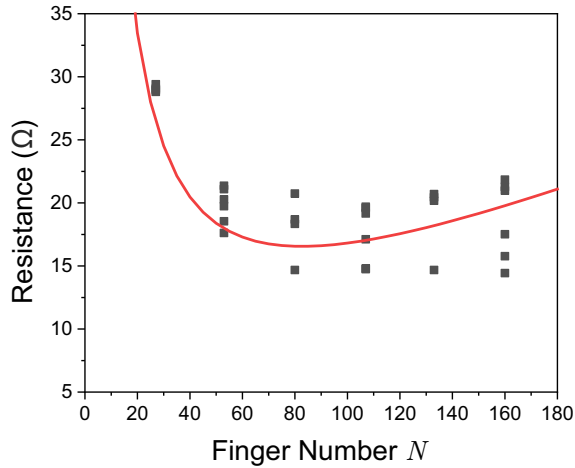

FIG. 2: Parasitic resistance $R_p$ as a function of number of finger pairs $N$ for several early prototype IDT devices (grey squares). The red line is from Eq. 9.

### A. Modeling of Parasitic Resistance $R_p$

The parasitic resistance $R_p$ can be thought of as having contributions from the long, wide side electrodes, and from the thin fingers themselves. The parasitic resistance of an inter-digitated electrode consisting of $N$ finger pairs is given by [4]:

$$R_p^{\text{idt}} = \frac{4}{3} \frac{l}{\Lambda t N} \rho, \tag{6}$$

where $l$ is the finger length, $\Lambda$ is the IDT pitch (equal to twice the finger width), $t$ is the metal thickness, and $\rho$ is the bulk resistivity of the Al metal. For a typical Al IDT used here,

$$R_p^{\text{idt}} = \frac{4}{3} \frac{0.24 \text{ mm}}{0.75 \text{ μm} \cdot 50 \text{ nm} \cdot 106} \left(2.7 \times 10^{-8} \text{ Ω} \cdot \text{m}\right) = 2.13 \text{ Ω}, \tag{7}$$

where we have used $\rho = 2.7 \times 10^{-8}$ Ω·m for aluminum. This resistance is much lower than the value of $R_p$ that we measure. If we consider the resistance of a single side bar, with designed width of 14 μm (cross-sectional area $A = t \times 14$ μm, and length $l^{\text{bar}} = 2\Lambda N + 40$ μm, this leads to a contribution

$$R_p^{\text{bar}} = \frac{\rho l^{\text{bar}}}{A} = \rho \frac{2\Lambda N + 40 \text{ μm}}{t \times 14 \text{ μm}}. \tag{8}$$

For our IDT, assuming $\rho$ is that of bulk aluminum, $R_p^{\text{bar}} = 7.7$ Ω, which is closer to the observed value of $R_p$. To first order, we can estimate the total parasitic resistance as resulting from these two contributions adding

in series:

$$R_{\mathrm{p}} \approx R_{\mathrm{p}}^{\mathrm{idt}} + R_{\mathrm{p}}^{\mathrm{bar}} = \frac{\rho}{t}\left(\frac{2\Lambda N + 40\ \mu\mathrm{m}}{14\ \mu\mathrm{m}} + \frac{8}{3}\frac{l}{\Lambda N}\right). \quad (9)$$

This simple model agrees quite well with results measured for a number of different IDT devices during early fabrication runs, as plotted in Fig. 2, where the red line corresponds to Eq. 9 with $\rho$ set to around twice the bulk resistivity of Al. (This may be a combination of increased resistivity in thin films and IDT fingers, as well as a structural factor due to the IDT geometry and electrode design.) These results suggest that $R_{\mathrm{p}}$ is currently limited by the width of the side electrode bars, and can be decreased by increasing this width and the film thickness $t$, as a path to increase the total electro-mechanical transduction efficiency of future IDT designs.

## II. ACOUSTO-OPTIC COUPLING STRENGTH AND DESIGN OPTIMIZATION

Acousto-optic coupling results from the photoelastic (or elasto-optic) effect, where an applied elastic strain field $S$ leads to a change in the material dielectric tensor $\varepsilon$ given by [1],

$$\Delta\varepsilon_{il} = \varepsilon_{ij} p_{jkmn} \varepsilon_{kl} S_{mn}, \quad (10)$$

where $p_{jkmn}$ are the elements of the material photoelastic tensor. In cubic media, such as silicon, there are only three independent photoelastic coefficients, which can be expressed using contracted notation with $11 \to 1, 22 \to 2, 33 \to 3$, and $23, 32 \to 4$, and $p_{13} = p_{12}$ by symmetry.

The SAWs excited through our experiments have strong in-plane displacement, resulting in an approximate elasto-optic refractive index change given by

$$\Delta n(x,y) \approx -\frac{1}{2} n_0^3 \left(p_{11} S_{xx} + p_{12} S_{yy}\right), \quad (11)$$

where $n_0$ is the equilibrium refractive index of the material. A time-harmonic strain field results in a spatiotemporally-varying index modulation, which can produce light scattering. Provided that $\Delta n(x,y)$ has the correct symmetry, this can lead to different forms of optical modulation, such as the form of inter-modal Bragg scattering studied in the latter half of the main article. In addition to phase-matching (correct overlap in the $z$ direction), the cross-sectional index modulation must have strong spatial overlap with the participating optical modes. Mathematically, this condition can be approximated through the acousto-optic coupling integral [1, 2, 5]:

$$g_{\mathrm{ao}} \propto \frac{\iint \left(n_0^2(\mathbf{r}_\perp) E_2^x(\mathbf{r}_\perp)\right)^* E_1^x(\mathbf{r}_\perp) \left(p_{11} S_{xx}(\mathbf{r}_\perp) + p_{12} S_{yy}(\mathbf{r}_\perp)\right) d\mathbf{r}_\perp}{\left(\iint |E_2^x(\mathbf{r}_\perp)|^2 d\mathbf{r}_\perp \iint |E_1^x(\mathbf{r}_\perp)|^2 d\mathbf{r}_\perp\right)^{1/2}}. \quad (12)$$

Here $E_1^x(\mathbf{r}_\perp)$ and $E_2^x(\mathbf{r}_\perp)$ are the transverse $x$-directed electric field profiles of the participating optical modes, and the acousto-optic modulation efficiency (power scattering rate) is $\propto |g_{\mathrm{ao}}|^2$. The integral is taken to be



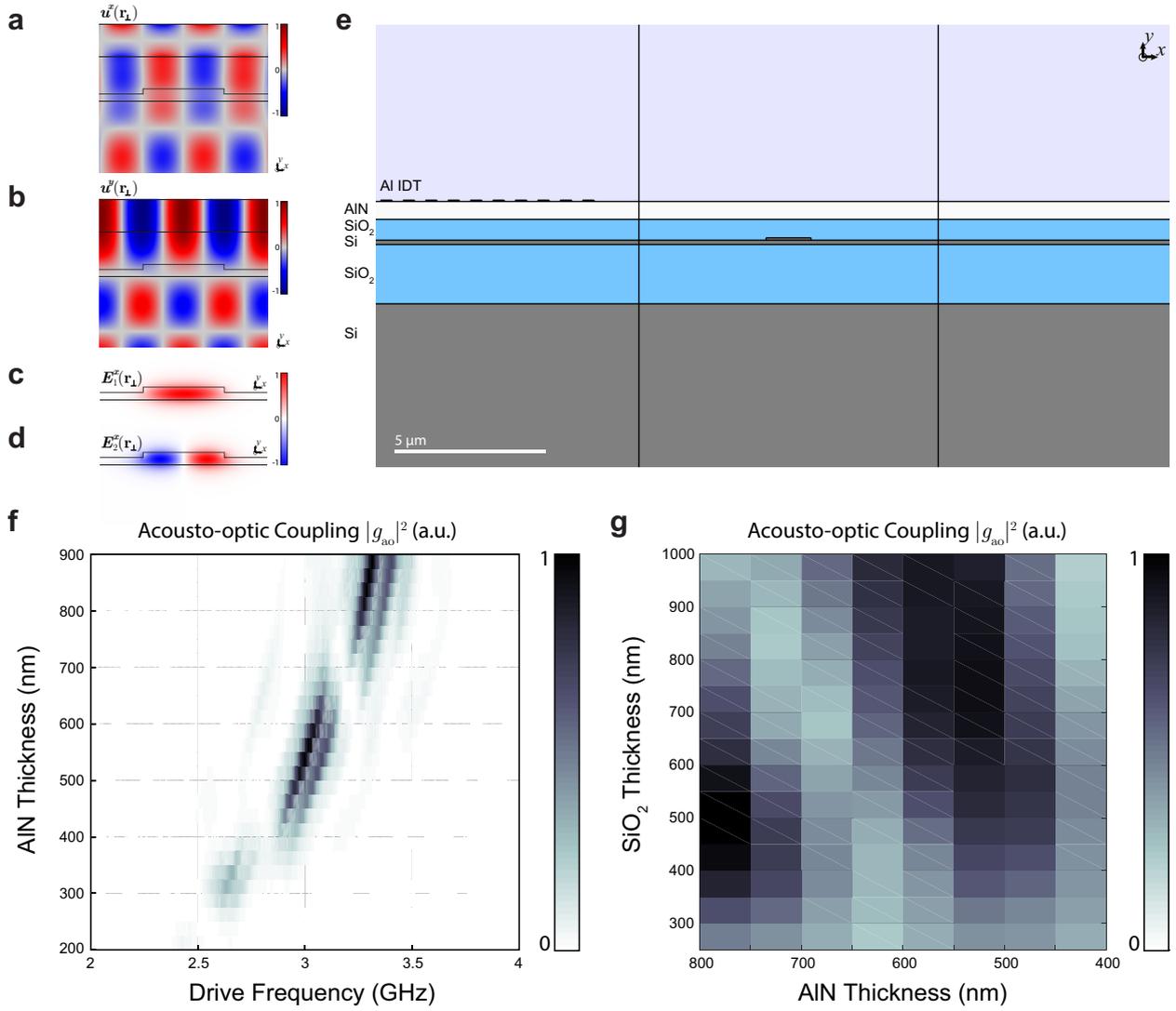

FIG. 3: Optimization of acousto-optic coupling strength. (a-b) depict simulated $x-$ and $y-$displacement profiles of a surface acoustic wave around a wavelength of $\Lambda_\text{ac} = 1.5$ μm (frequency $\Omega = 3.1$ GHz) used to mediate inter-modal acousto-optic scattering. (c) and (d) depict the $x-$directed electric field profiles of symmetric and anti-symmetric optical modes guided by a silicon ridge waveguide of $w = 1.5$ μm, which are coupled by the elastic wave in (a-b). (e) depicts a portion of the geometry of a model cross-sectional device region used to simulate acousto-optic interactions. The domain extends further in the $-x$ direction for a total of 18 IDT finger pairs. (f) plots simulated acousto-optic coupling strength $|g_\text{ao}|^2$ as a function of drive frequency and AlN layer thickness for one device geometry. (g) plots the maximum acousto-optic coupling strength as a function of AlN and SiO$_2$ layer thicknesses for the same waveguide geometry.

over the transverse cross-section of the optical waveguide. In our structure, the primary contribution is from the silicon waveguide core, which tightly confines the optical modes and has photoelastic coefficients $(p_{11}, p_{12}, p_{44}) = (-0.09, 0.017, -0.051)$ [6]. Though surface waves typically have strong displacement in the $y$ direction (normal to the material surface or interface), this results in a relatively small elasto-optic modulation due to silicon's small $p_{12}$ coefficient. As plotted in Fig. 3a-b, the types of SAWs used to produce modulation in our structures have substantial displacement in the $x$ direction, allowing this interaction to access silicon's large $p_{11}$ coefficient.



For intra-modal phase modulation, simulations show that $g_{ao}$ is maximized when the acoustic wavelength $\Lambda_{ac} \approx 2w$, where $w$ is the optical waveguide width. As plotted in Fig. 3a-d, for inter-modal modulation we choose elastic waves with $\Lambda_{ac} \approx w$, so that the SAW profile optimizes optical scattering between symmetric and anti-symmetric optical modes guided in the waveguide core.

For optimal overlap, it is also essential that the SAW has large displacement (strain) in the waveguide core, which is affected by the film thicknesses of the material stack. Finally, thickness of the piezoelectric AlN layer should be optimized to produce efficient electro-mechanical transduction. These parameters are varied through finite element simulations using the commercially-available software package COMSOL, with simulation geometry and typical results shown in Fig. 3e-g. Based on these results, which show optimal coupling around film thicknesses $t_{AlN} = 550$ nm and $t_{SiO2} > 600$ nm, we chose values of $t_{AlN} = 600$ nm and $t_{SiO2} = 700$ nm to optimize coupling for devices with $w = 1.5$ μm.

Because SAWs tend to only penetrate to a depth that is a few multiples of $\Lambda_{ac}$, this same film stack is not optimal for acousto-optic coupling at high frequencies (small SAW wavelengths). As a result, observed acousto-optic modulation efficiencies decrease above $\approx 4$ GHz, as shown in Fig. 2 of the main article. To implement devices with robust operation in the $\sim$5-10+ GHz range, the thicknesses of the films above the optical waveguide layer should be decreased.

### III. PHASE-MATCHING AND DESIGN FOR INTER-MODAL MODULATORS

The inter-modal modulation process is mediated by an incident SAW with a well-defined axial wavevector $q$, with phase-matching conditions of the form:

$$k_-(\omega_p + \Omega) = k_+(\omega_p) + q. \tag{13}$$

where $k_+(\omega)$ and $k_-(\omega)$ refer to the dispersion relations for the symmetric and anti-symmetric optical modes. This process is plotted in Fig. 4b, where the SAW with frequency $\Omega$ and wavevector $q$ scatters light between an initial optical state, represented by an open circle at optical frequency $\omega_p$ and wavevector $k_+(\omega_p)$, and a final optical state, represented by a filled circle at optical frequency $\omega_p + \Omega$ and wavevector $k_-(\omega_p + \Omega)$. This phase-matching condition can be rewritten as:

$$\frac{(\omega_p + \Omega) \times n_{p,-}(\omega_p + \Omega) - \omega_p \times n_{p,+}(\omega_p)}{c} = q, \tag{14}$$

where the muliplication symbols ($\times$) were added for clarity. Using the approximations that $|n_{p,-} - n_{p,+}|\omega_p \gg \Omega \times n_{p,+}$, and that $n_{p,-}$ does not change dramatically over a frequency detuning $\Omega$, this expression can be rewritten as

$$\frac{2\pi}{\lambda_p}(n_{p,-}(\lambda_p) - n_{p,+}(\lambda_p)) = q. \tag{15}$$

By simulating $n_{p,-}$ and $n_{p,+}$ as a function of wavelength for a given waveguide structure, this equation serves as a design rule to specify the target value for $q$.

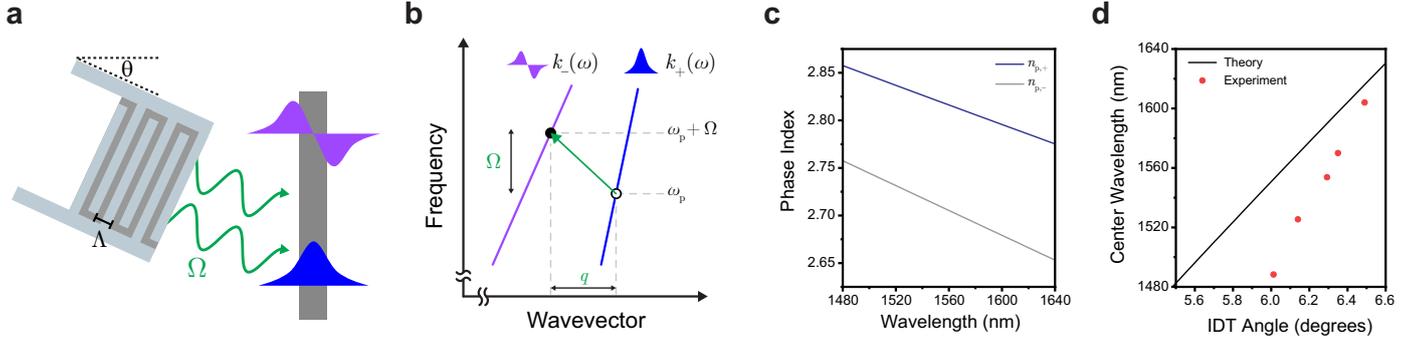

FIG. 4: IDT design and inter-modal phase matching. (a) For the purposes of phase-matching, the design of the inter-digitated transducer is defined by a inter-digital pitch $\Lambda$ and angle $\theta$, with respect to the direction normal to the optical axis. This transducer emits a SAW with frequency $\Omega$ and wavevector $q$ set by these parameters. (b) Through an inter-modal acousto-optic scattering process, this SAW mediates transitions between points on two optical dispersion bands at two optical frequencies separated by $\Omega$. (c) Simulated values of phase indices $n_{p,+}(\lambda)$ and $n_{p,-}(\lambda)$ for a device with $w = 1500$ nm. (d) Simulated values (line) and measured (dots) of center modulation wavelength $\lambda_0$ vs. designed IDT angle $\theta$ for a device with $w = 1500$ nm.

The SAW is generated electro-mechanically using an IDT with pitch $\Lambda$ defined by the inter-digital spacing as plotted in Fig. 4a. This transducer couples to SAWs with frequencies $\Omega^i = v_s^i/\Lambda$, where $v_s^i$ is the acoustic propagation velocity of mode $i$.

The transducer is oriented with an angle $\theta$ with respect to normal of the optical axis. This angle directly sets the axial wavevector $q$, and relates the transducer pitch $\Lambda$ to the transverse acoustic wavelength (from the perspective of the optical waveguide) $\Lambda_{ac}$ by [7]:

$$\Lambda = \frac{1}{2}\Lambda_{ac}\cos(\theta); \tag{16}$$

$$q = (2\pi/\Lambda_{ac})\tan(\theta) = (\pi/\Lambda)\sin(\theta). \tag{17}$$

Combined with the overlap rule that $\Lambda_{ac} \approx w$ for efficient inter-modal scattering, where $w$ is the waveguide width, Eqs. 15 and 17 lead to the expression

$$\tan(\theta) = \frac{w}{\lambda_p}\left(n_{p,-}(\lambda_p) - n_{p,+}(\lambda_p)\right), \tag{18}$$

which specifies the angle $\theta$ necessary to mediate inter-modal scattering at wavelength $\lambda_p$. Fig. 4c plots the simulated values of $n_{p,+}$ and $n_{p,-}$ as a function of optical wavelength. Fig. 4d plots the corresponding calculated and measured values of $\lambda_0$, the central wavelength for which phase-matching is exactly satisfied, as a function of designed IDT angle $\theta$, for several devices. Reasonable agreement is achieved, in particular at longer wavelengths. At shorter wavelengths, the simulated values of the phase indices (and their difference) may differ from the actual values due to uncertainties in device geometry.

## IV. OPERATING BANDWIDTH OF INTER-MODAL AND NON-RECIPROCAL MODULATORS

For a given SAW wavevector $q$, Eq. 13 is exactly satisfied at a single optical frequency $\omega_0$ (wavelength $\lambda_0$). However, as the operating frequency $\omega_p$ is detuned from $\omega_0$, the acousto-optic scattering process is no longer precisely phase-matched, with an approximate wavevector mismatch $\Delta q_{pm}$ found by linearizing Eq. 13 around $\omega_p = \omega_0$ as [8]:

$$\Delta q_{pm} = \frac{\partial k_-}{\partial \omega}(\omega - \omega_p) - \frac{\partial k_+}{\partial \omega}(\omega - \omega_p) = \frac{n_{g,-} - n_{g,+}}{c}\Delta\omega, \quad (19)$$

where $n_{g,+}$ and $n_{g,-}$ are the group velocities of the symmetric and anti-symmetric modes, and $\Delta\omega = \omega_p - \omega_0$.

The total modulated optical power $P_{as}$ is related to the integrated phase mismatch over a propagation length $L$ as given by the integral [8]:

$$P_{as}(L) \propto \left| \int_0^L g_{ao}(z) e^{i\Delta q_{pm} z} dz \right|^2. \quad (20)$$

Assuming that the acousto-optic coupling rate $g_{ao}(z)$ is constant along the propagation length $z$, this integral can be evaluated as:

$$P_{as}(L) \propto |g_{ao}|^2 \operatorname{sinc}^2\left(\Delta q_{pm} L/2\right). \quad (21)$$

This $\operatorname{sinc}^2$ response envelope is equal to $1/2$ when $\Delta q_{pm} L/2 = 1.39$, resulting in a full-width at half-maximum (FWHM) operation bandwidth $\Delta\omega_{FWHM}$ given by

$$\Delta\omega_{FWHM} = \frac{4 \cdot 1.39 c}{L} \frac{1}{|n_{g,+} - n_{g,-}|}, \quad (22)$$

where $c$ is the speed of light. This bandwidth can be expressed in terms of wavelength as

$$\Delta\lambda_{FWHM} = \frac{2 \cdot 1.39}{\pi |n_{g,-} - n_{g,+}|} \frac{\lambda_0^2}{L}. \quad (23)$$

For a given device length, the operating bandwidth can be increased by dispersion engineering that tailors the group indices of the propagating modes to be nearly equal (see Ref. [8] for details).

### A. Wavelength Response Engineering and Conditions for Non-reciprocity

As discussed in the main article, an inter-modal modulator induces a non-reciprocal wavevector mismatch $\Delta q_{nr}$ between forward- and backward-propagating waves, where

$$\Delta q_{nr} \approx \frac{\Omega}{c}(n_{g,-} + n_{g,+}). \quad (24)$$

This term implies that the center operation frequency $\omega_0$ takes on different values $\omega_f$ and $\omega_b$ for forward- and backward-propagating optical waves, respectively, given by

$$\omega_f - \omega_b = \frac{n_{g,-} + n_{g,+}}{n_{g,-} - n_{g,+}}\Omega. \quad (25)$$

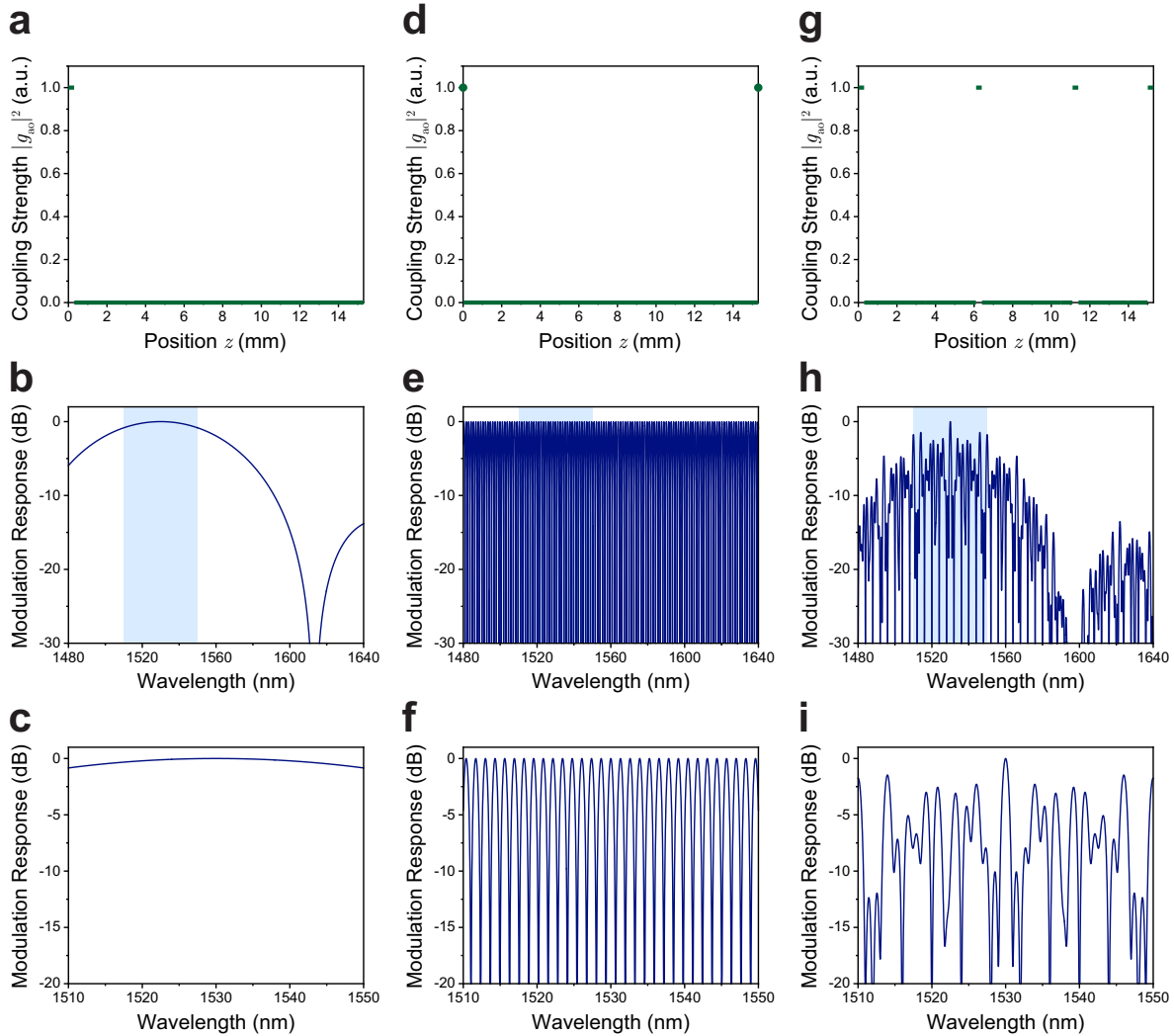

FIG. 5: Wavelength-dependent response as a function of device design. (a-c) depict the design and response of a device with a single 240 µm acousto-optic interaction region. (a) plots the coupling strength as a function of position $z$, whereas the resulting modulation response is shown in (b) (wideband) and (c) (zoomed-in view of the shaded region). (d-f) depict the design and response of a device with delta-function coupling at two points separated by $L = 15.3$ mm. (d) plots this coupling strength as a function of position $z$, while the resulting modulation response is shown in (e) (wideband) and (f) (zoomed-in). (g-i) depict the design and response of a device similar to the serpentine modulator studied in the article, where coupling occurs over four 240-µm-long regions spaced along a total length of $L = 15.3$ mm. (g) plots this coupling strength as a function of position $z$, while the resulting modulation response is shown in (h) (wideband) and (i) (zoomed-in).

Expressed simply, the condition for significant non-reciprocal modulation contrast between forward and backward propagating waves, necessary to build a high-extinction non-reciprocal modulator is:

$$|\omega_\text{f} - \omega_\text{b}| \gg \Delta\omega_\text{FWHM}, \tag{26}$$

which places a constraint on device length to support strong non-reciprocity:

$$L \gg \frac{4 \cdot 1.39}{n_{\text{g},-} + n_{\text{g},+}} \frac{c}{\Omega}. \tag{27}$$

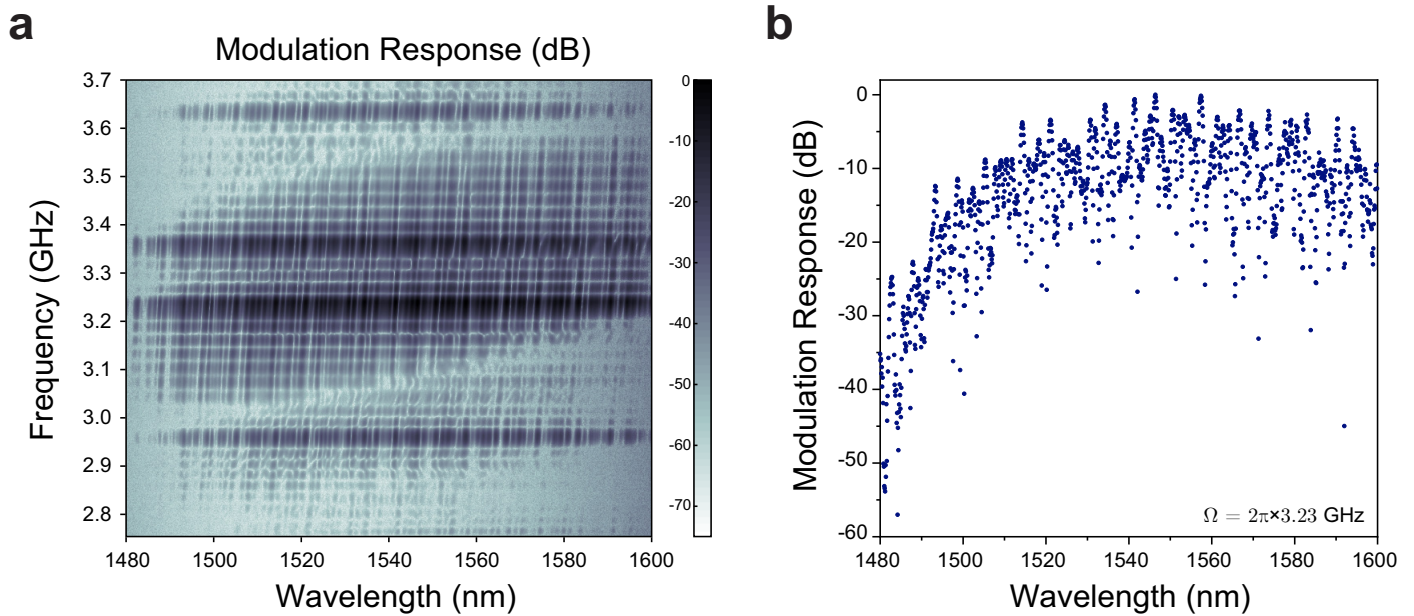

FIG. 6: Wideband experimental measurements of modulation response in a serpentine modulator device. (a) plots the observed modulation response as a function of drive frequency and wavelength. (b) plots a single slice at a drive frequency of $\Omega/2\pi = 3.23$ GHz.

For one of our fabricated optical waveguides, with measured $n_{g,-} = 3.84$, $n_{g,+} = 3.72$, and $\Omega = 2\pi \times 3.11$ GHz, this condition is $L \gg 1.1$ cm. However, due to the relatively small IDT apertures used, our acousto-optic interaction lengths are much shorter than this length ($L_{ao} = 0.24 - 0.96$ mm). Therefore, to artificially achieve a sufficiently narrow operation bandwidth for non-reciprocal modulation, we can distribute this modulation over a much longer optical propagation length through the use of the serpentine modulator design.

We numerically study this behavior using Eq. 20 with different distributions of $g_{ao}(z)$ as shown in Fig. 5. A constant modulation strength from $z = 0$ to $z = 0.24$ mm, similar to a single inter-modal modulator device, is depicted in Fig. 5a-c, leading to the expected broadband $\text{sinc}^2$ response from Eq. 21.

As a point of comparison, we consider a device where the coupling $g_{ao}(z) = 0$ except at $z = 0$ and $z = 15.3$ mm (a double-$\delta$-function modulation profile.) The resulting modulation strength as a function of frequency detuning can be calculated from Eq. 20 as

$$P_{as}(L) \propto \cos^2\left(\Delta q_{pm} L/2\right). \tag{28}$$

This response is studied in Fig. 5d-f. Even with modulation only at two points, provided there is a sufficient time delay, we can engineer a narrowband and periodic optical response necessary for non-reciprocity.

Finally, we consider the case of the serpentine modulator studied in the text, where $g_{ao}(z)$ is nonzero for four 0.24-mm-long segments distributed along the total device length $L_{tot} = 15.3$ mm. As depicted in Fig. 5g-i, this type of device design leads to a semi-periodic non-reciprocal response with a wider envelope corresponding to the shorter constituent segment lengths. As more segments or longer lengths are added, the wavelength response becomes similar to the narrowband $\text{sinc}^2$ response of a single modulator of length $L_{tot}$.

Experimental data are plotted for a single serpentine modulator of $\Lambda = 721$ nm, $\theta = 6.28°$, $w = 1.45$ and μm. This device has measured center wavelength $\lambda_\text{f} = 1544$ nm; its modulation response is plotted as a function of $\Omega$ and $\lambda_\text{p}$ in Fig. 6a, and at a fixed frequency $\Omega/2\pi = 3.23$ GHz, in Fig. 6b. The steep vertical slants in Fig. 6a correspond to the phase-matching condition from Eq. 14.

## V. ANTI-STOKES AND STOKES INTER-MODAL MODULATION

In this article, we studied inter-modal acousto-optic modulation and mode conversion through anti-Stokes (blue-shifting) scattering processes. However, all of the main results can be reproduced with similar behavior using a Stokes (red-shifting) scattering process, as studied in Fig. 7. The incident SAW can be seen as mediating optical transitions between two points on distinct optical dispersion branches (optical waveguide modes). For a backward-propagating acoustic wave, light injected into the symmetric waveguide mode is anti-Stokes ($+\Omega$)-shifted whereas light injected into the anti-symmetric mode is Stokes ($-\Omega$)-shifted.

Forward and backward modulation response data within the serpentine modulator device studied in the main article are plotted for both anti-Stokes (Fig. 7c) and Stokes (Fig. 7f) scattering. The wavelength-dependence of the modulation efficiency and the non-reciprocal response are similar for both processes, though with a slight blue-shift ($\sim \Omega$) in the Stokes modulation response due to phase-matching. The output light is red-shifted relative to the input through the Stokes process, which may be preferable for some applications. This equivalence also highlights the need for low-crosstalk mode multiplexers, since light multiplexed into the unwanted mode produces scattering to the opposite (potentially unwanted) frequency sideband.

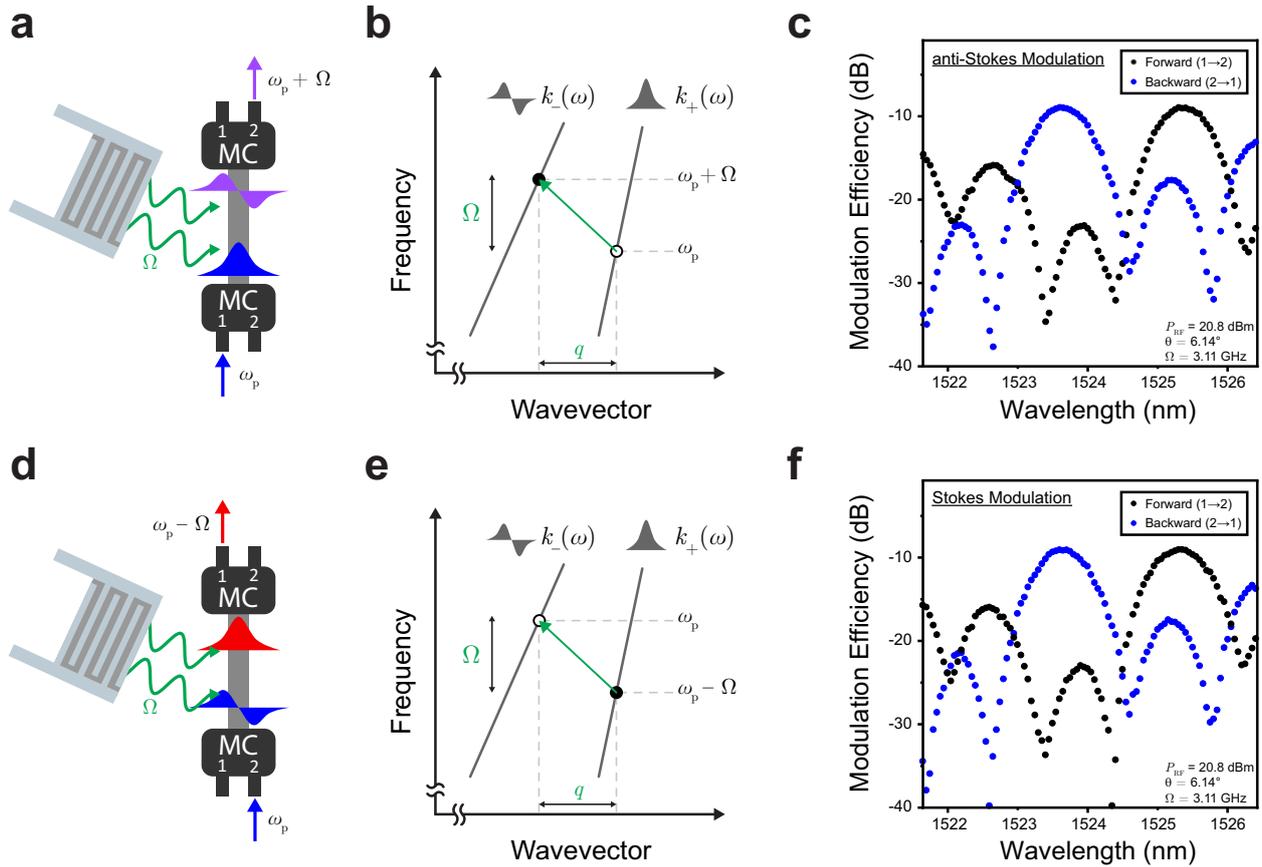

FIG. 7: Stokes and anti-Stokes inter-modal modulation. (a-c) depict the device operation scheme studied in the main article, where inter-modal acousto-optic modulation is implemented through an anti-Stokes (blue-shifting) scattering process. (a) depicts the path of optical waves through this process. Probe light at frequency $\omega_p$ is injected into port 1 of a mode multiplexer and coupled into the symmetric waveguide mode. As it propagates through the silicon waveguide, it is scattered by an impinging elastic wave to $\omega_p + \Omega$ and into the anti-symmetric waveguide mode, before exiting the device through port 2 of a second, structurally-identical mode multiplexer. (b) plots phase-matching for this process, where the incident elastic wave with (frequency, wavevector) $= (\Omega, q)$ mediates transition between initial (open circle) and final (filled circle) optical states on two dispersion bands. (c) plots modulation through this process within the serpentine modulator device studied in the main article, showing a characteristic non-reciprocal response. (d-f) depict alternate operation through a Stokes (red-shifting) scattering process. (d) depicts the path of optical waves through this process. Probe light at frequency $\omega_p$ is injected into the anti-symmetric waveguide mode through port 2 of a mode multiplexer. As it propagates through the waveguide, it is scattered by the incident elastic wave to $\omega_p - \Omega$ and into the symmetric waveguide mode, before exiting the device through port 1 of an identical mode multiplexer. (e) plots phase-matching for this process, which is similar to (b) except that the initial and final states are switched. (f) plots the modulation response of the same serpentine modulator device through the Stokes process, showing a similar behavior with wavelength.



\end{document}